%% file: main_archive_Rev1_resubmission.tex
\documentclass[12pt]{article}

\PassOptionsToPackage{table}{xcolor}
\usepackage{moreverb,url}
\usepackage{multirow,multicol,makecell,booktabs}
\usepackage{subcaption}
\usepackage{caption}
\usepackage{bm}
\usepackage{amsmath}
\usepackage{colortbl}
\usepackage{xcolor}
\usepackage{array}
\usepackage{amssymb} 

\usepackage{graphics}
\usepackage{graphicx}
\usepackage{rotating}
\usepackage{tabularx}
\usepackage{hyperref}
\usepackage[numbers]{natbib} 
\usepackage{booktabs}
\usepackage{xr}
\usepackage{amsthm} 
\newtheorem{proposition}{Proposition} 

\newcommand\BibTeX{{\rmfamily B\kern-.05em \textsc{i\kern-.025em b}\kern-.08em
T\kern-.1667em\lower.7ex\hbox{E}\kern-.125emX}}

\usepackage[margin=1in]{geometry}
\begin{document}

\title{Moving toward best practice when using propensity score weighting in survey observational studies}

\author{
  Yukang Zeng$^{1,2,3}$, \ 
  Fan Li$^{1,3,4}$, \ 
  Guangyu Tong$^{1,2,3,4}$ \\
  \\
  {\small $^1$ Department of Biostatistics, Yale School of Public Health, New Haven, CT, USA} \\
  {\small $^2$ Cardiovascular Medicine Analytics Center, Yale School of Medicine, New Haven, CT, USA} \\
  {\small $^3$ Section of Cardiovascular Medicine, Department of Internal Medicine,} \\
  {\small \quad Yale School of Medicine, New Haven, CT, USA} \\
  {\small $^4$ Center for Methods in Implementation and Prevention Science,} \\
  {\small \quad Yale School of Public Health, New Haven, CT, USA} \\
  \\
  {\small Corresponding Author: Guangyu Tong, PhD} \\
  {\small Email: \texttt{guangyu.tong@yale.edu}}
}

\date{\vspace{0.3em} \today}

\maketitle

\begin{abstract}
    Propensity score weighting is a common method for estimating treatment effects with observational data, by addressing confounding due to measured baseline covariates. However, when the observational data sample is drawn based on a survey, the existing literature does not reach a consensus on the optimal use of survey weights for population-level causal inference. Under the balancing weights framework, we provide a unified solution for incorporating survey weights and derive a set of weighting and augmented weighting estimators for different target populations, including the combined, treated, controlled, and overlap populations. We also develop closed-form sandwich variance estimators for each estimator via the theory of M-estimators. Through an extensive series of simulation studies, we examined the performance of our estimators and compared the results to those of alternative methods. We carried out two case studies to illustrate the application of the different propensity score methods with complex survey data. We concluded with a discussion of our findings and provided some practical recommendations for propensity score weighting analysis of survey observational data.
\end{abstract}

\vspace*{1em}

\section{Introduction}\label{sec1}

The use of observational data to estimate the causal effects is prevalent in medical research, particularly when randomized trials are unavailable. In many cases, researchers resort to observational data garnered through surveys with complex sampling techniques. These techniques are designed to ensure data from a sample can accurately represent the external target population, and include features such as multistage sampling and deliberate oversampling of less representative subgroups.\cite{LaVange2001Applying} The survey weights computed based on the design are integral to recovering the population features so that the differentiated selection probabilities and non-responses are appropriately accounted for. There is consensus on the importance of incorporating the survey design into the analytical process to derive causal effect estimates that are applicable to the target population.\cite{Zanutto2006,DuGoff2014,Ridgeway2015}

Survey weights also play a crucial role in ensuring the accuracy and representativeness of the propensity score method, which is among the most commonly used tool for causal inference using observational data.\cite{ImaiKingStuart2008} In the absence of randomization, systematic differences in baseline characteristics usually exist between exposed and unexposed individuals, leading to the issue of confounding. The propensity score methods act on the treatment assignment mechanism with measured baseline covariates,\cite{Rosenbaum1983} and with estimated propensity scores, differences in the observed confounders across groups can be reduced so that the estimated treatment effect will no longer be attributable to pre-existing group differences. In practice, implementing propensity score methods involve a two-step procedure, including propensity score estimation and the subsequent treatment effect estimation. The latter step leverages the estimated propensity score through stratification,\cite{Lunceford2004} matching,\cite{Stuart2010} or weighting.\cite{Li2018}

The appropriate use of survey weights in stratification to obtain population causal effect estimates has been previously addressed, with a recommendation that survey weights should be used when estimating treatment effects within strata but not in the estimation of the propensity scores. \cite{ZanuttoLuHornik2005,Zanutto2006} DuGoff et al.\cite{DuGoff2014} later suggested that survey weights can carry unique design features that are not captured by observed covariates, and therefore should be included as an additional covariate in the propensity score model to enhance covariate balance and address potential bias due to unmeasured covariates. Similar recommendations have also been extended to propensity score matching and weighting, where the estimation of propensity score can incorporate survey weights as an additional covariate; but in those contexts, whether to perform survey-weighted analysis in the outcome model depends on the goal of the analysis (i.e., whether a sample or the underlying population defines the target of inference).\cite{DuGoff2014} Austin et al.\cite{AustinJembereChiu2018} illustrated through simulations that including survey weight as a covariate in the propensity score model showed no clear advantage over a survey-weighted propensity score model. Neither approach can consistently outperform the other regarding covariate balance and bias in the treatment effect estimation across various scenarios. Dong et al. \cite{DongStuartLenisNguyen2020} evaluated multiple strategies for incorporating survey weights into both propensity score and outcome models, recommending their use at both stages, although they observed that estimates from different strategies did not yield apparent differences in their data application. Lenis et al.\cite{Lenis2019} found through simulations that incorporating survey weights in the propensity score model does not necessarily improve the estimation of population-level treatment effect using matching, especially when sufficient population-level balance in confounders can already be achieved. They also showed through simulations that incorporating survey weights in propensity score estimation can reduce bias when the missingness in outcome is correlated with survey weights.\cite{Lenis2019}

Regarding propensity score weighting specifically, Ridgeway et al.\cite{Ridgeway2015} provided a theoretical justification for the use of a survey-weighted propensity score model in obtaining a consistent estimator of the population-level average treatment effect. They proved that using survey weights in both stages leads to consistent and robust population average treatment effect estimators under various generating mechanisms. Salerno et al. \cite{salerno2025whatsweight} recently explored a complex survey mechanism in health disparities research, where survey weights depend on the group variable of interest under comparison, such as race. They further revealed that whether the propensity score should be survey-weighted depends on whether the available survey weights are tied to comparison variables such as treatment or exposure. Whereas Ridgeway et al.\cite{Ridgeway2015} derived the results and performed simulations for the continuous outcome scenario, Yang et al.\cite{Yang2023} extended the results to a binary outcome and evaluated the performance of different ways to incorporate survey weights in the propensity score model and the outcome model. Both Ridgeway et al.\cite{Ridgeway2015} and Yang et al.\cite{Yang2023} considered the potentially differentiated practice of incorporating survey weights when estimating the average treatment effect (ATE) and the average treatment effect of the treated (ATT).

In general, the emerging message from the existing literature is that the use of survey weights at the propensity score model stage and the outcome model stage can potentially be beneficial in reducing the bias of population-level inference. However, a consensus on the best practice and the underlying justifications for using survey weights to target different causal estimands of interest have not been fully elucidated. This paper aims to contribute to the literature on propensity score weighting in several ways. First, the target estimands in the prior development on survey observational data methods were limited to ATE and ATT, and have not covered other weighted average treatment effect estimands.\cite{Li2018} We address this by supplying a unified framework that incorporates survey weights into the general framework of balancing weights.\cite{Li2018} Second, as a further extension, we detail the use of survey weights for augmented weighting estimators under the balancing weights framework, which can potentially improve efficiency over weighting alone by leveraging information from a posited outcome regression function. Third, we developed closed-form variance estimators that accommodate survey weights in each estimator we considered for computationally efficient inference. Lastly, given that no consensus has been reached on the best use of survey weights in propensity score weighting analysis under various real-world scenarios (such as varying levels of covariate overlap and model misspecification), we conduct extensive simulations to shed light on the operating characteristics of these methods and provide practical recommendations.  

The rest of this paper is organized as follows. Section \ref{sec2} defines the notation and assumptions for the potential outcomes framework in the context of survey observational data. Under the balancing weight framework, this section also presents the weighting estimator and discusses the role of survey weights in addition to balancing weights. 
In Section \ref{sec3}, we present three augmented estimators incorporating survey weights and further provide the the associated sandwich variance structure for large-sample inference. In Section \ref{sec4}, we conduct simulations to examine the performance of different estimators incorporating propensity score weights for targeting different estimands. Section \ref{sec5} compares the performance of the different estimation procedures in two real data examples, and Section \ref{sec6} concludes.  

\section{Methods for incorporating survey weights into propensity score weighting}\label{sec2}

\subsection{Definitions and assumptions}\label{sec2.1}

Consider a finite sample of size \(n\) selected from a original population of size \(N\). Each individual in the sample is indexed by $i$,  where $i = 1,\dots, n$. We consider a $p$-dimensional vector for pre-treatment covariates for each individual \(\mathbf{X}_i\), and a binary exposure variable \(Z_i = z\) that indicates treatment (\(z = 1\)) or control (\(z = 0\)) group affiliation. Under the Stable Unit Treatment Value Assumption (SUTVA), the potential outcomes \(Y_i(z)\) for \(z \in \{0, 1\}\) are assumed to be well-defined, and the observed outcome for each individual is \(Y_i = Z_i Y_i(1) + (1 - Z_i) Y_i(0)\).
For estimating population effects using our sample, we define the survey indicator \(S_i\) where \(S_i = 1\) denotes selection into the sample and \(S_i = 0\) otherwise. Following Salerno et al.,\cite{salerno2025whatsweight} we distinguish two sampling regimes based on when sampling occurs relative to treatment assignment. In a \emph{retrospective} design, sampling occurs after treatment has been assigned in the population, and the sampling mechanism may depend on both treatment status and covariates. This yields treatment-specific sampling probabilities 
$$p_{z}(\mathbf{x}_i) \equiv P(S_i=1 \mid Z_i=z,\mathbf{X}_i=\mathbf{x}_i)~~~ \text{for}~~~ z\in\{0,1\},$$ 
with the positivity constraint \(0<p_z(\mathbf{X}_i)<1\). In a \emph{prospective} design, sampling occurs before treatment assignment, so the sampling mechanism depends only on covariates with sampling probability 
$$p(\mathbf{x}_i) \equiv P(S_i=1 \mid \mathbf{X}_i=\mathbf{x}_i),$$
which satisfies \(0<p(\mathbf{X}_i)<1\). Under the prospective sampling, \(S \perp Z \mid \mathbf{X}\) implies \(p_1(\mathbf{x})=p_0(\mathbf{x})=p(\mathbf{x})\). We further define the population-level propensity score \(e_{\text{sp}}(\mathbf{x}_i) \equiv P(Z_i = 1 \mid \mathbf{X}_i = \mathbf{x}_i)\) and the sample-level propensity score \(e_{\text{fp}}(\mathbf{x}_i) \equiv P(Z_i = 1 \mid \mathbf{X}_i = \mathbf{x}_i, S_i = 1)\), both satisfying positivity constraints \(0<e_{\text{sp}}(\mathbf{X}_i)<1\) and \(0<e_{\text{fp}}(\mathbf{X}_i)<1\) on the covariate support.

In a retrospective design, we assume conditional independence (or unconfoundedness) for the treatment assignment, \(Y(z) \perp Z \mid \mathbf{X}\), and for the post-exposure sampling process, \(Y(z) \perp S \mid Z, \mathbf{X}\). These conditions imply that confounding bias arises solely from \(\mathbf{X}\), but selection bias can be caused by \(Z\) and \(\mathbf{X}\). 
In contrast, a prospective design requires slightly different assumptions. Under noninformative prospective sampling, we assume that potential outcomes are independent of sampling given only covariates, \(Y(z) \perp S \mid \mathbf{X}\), and that sampling and treatment are conditionally independent, \(S \perp Z \mid \mathbf{X}\). Furthermore, we assume that treatment assignment is unconfounded within the sample such that \(Y(z) \perp Z \mid S=1, \mathbf{X}\). These conditions imply that both selection bias and subsequent treatment assignment bias arise solely through \(\mathbf{X}\), simplifying the identification strategy.

Under the retrospective designs, the relationship between population and sample propensity scores follows directly from Bayes' rule:
\begin{equation}\label{eq:efp-retro-ps}
e_{\text{fp}}(\mathbf{x})
=\frac{p_1(\mathbf{x})\,e_{\text{sp}}(\mathbf{x})}{p_1(\mathbf{x})\,e_{\text{sp}}(\mathbf{x})+p_0(\mathbf{x})\{1-e_{\text{sp}}(\mathbf{x})\}}
=\operatorname{expit}\!\left(\operatorname{logit}e_{\text{sp}}(\mathbf{x})+\log\displaystyle\frac{p_1(\mathbf{x})}{p_0(\mathbf{x})}\right).
\end{equation}
This identity reveals how treatment-dependent sampling tilts the population propensity scores: the log conditional sampling ratio \(\log\{p_1(\mathbf{x})/p_0(\mathbf{x})\}\) shifts the population-level propensity score to produce the sample-level propensity score. Large discrepancies in sampling rates, either \(p_1(\mathbf{x}) \gg p_0(\mathbf{x})\) or \(p_1(\mathbf{x}) \ll p_0(\mathbf{x})\), can drive \(e_{\text{fp}}(\mathbf{x})\) toward the boundaries, threatening sample-level positivity even when population-level positivity holds. In contrast, under prospective designs, we have \(p_1(\mathbf{x})=p_0(\mathbf{x})=p(\mathbf{x})\), so Equation \eqref{eq:efp-retro-ps} reduces to \(e_{\text{fp}}(\mathbf{x}) \equiv e_{\text{sp}}(\mathbf{x})\). This distinction clarifies when survey-weighted propensity scores are necessary for consistent inference: under retrospective (informative) sampling, survey-weighted estimation is required to recover $e_{\text{sp}}(\mathbf{x})$; under prospective (non-informative) sampling where $e_{\text{fp}}(\mathbf{x}) = e_{\text{sp}}(\mathbf{x})$, unweighted estimation is theoretically valid.

The above relationship carries implications for certain estimators we will introduce later. To complete the definition of notation, we define \(r_{z}(\mathbf{x}) = p(\mathbf{x})/p_z(\mathbf{x})\) for \(z \in \{0,1\}\). From Equation \eqref{eq:efp-retro-ps} and \(p(\mathbf{x}) = e_{\text{sp}}(\mathbf{x})p_1(\mathbf{x}) + \{1-e_{\text{sp}}(\mathbf{x})\}p_0(\mathbf{x})\), we arrive at \(r_1(\mathbf{x}) = e_{\text{sp}}(\mathbf{x})/e_{\text{fp}}(\mathbf{x})\) and \(r_0(\mathbf{x}) = \{1-e_{\text{sp}}(\mathbf{x})\}/\{1-e_{\text{fp}}(\mathbf{x})\}\), which together yield \(p_1(\mathbf{x})/p_0(\mathbf{x}) = r_0(\mathbf{x})/r_1(\mathbf{x})\). Here \(r_1(\mathbf{x}) > 1\) indicates undersampling of treated units relative to their population prevalence, while \(r_1(\mathbf{x}) < 1\) indicates oversampling. The relationship \(p(\mathbf{x}) = r_z(\mathbf{x})p_z(\mathbf{x})\) provides a bridge for estimating marginal sampling probabilities when only treatment-specific ones are available from survey data, under retrospective designs. As Salerno et al.\cite{salerno2025whatsweight} highlighted, the distinction between prospective and retrospective studies corresponds to two different factorizations of \(P(S=1, Z=z \mid \mathbf{X})\): one conditional on covariates alone versus one conditional on both treatment and covariates. In this paper, we primarily focus on the retrospective design, but provide the counterparts of our methods under prospective design in Web Appendix A.6.

We define the conditional average treatment effect as \(\tau(\mathbf{x}) = E[Y(1) - Y(0) \mid \mathbf{X} = \mathbf{x}]\), which is the building block for marginal estimands depending on the covariate distribution. 
Although prior work has primarily focused on the population average treatment effect (PATE) and the population average treatment effect of the treated (PATT),\cite{Ridgeway2015,Yang2023,salerno2025whatsweight} we follow the framework of balancing weights \cite{Li2018} that characterizes a general class of weighted distributions of covariates with an extension to survey observational studies. 
Specifically, the density of the target population is represented as \(g(\mathbf{X}) = f(\mathbf{X})  h(\mathbf{X})\), where \(f(\mathbf{X})\) signifies the marginal density of covariates of the original external population, and \(h(\mathbf{X})\), known as the tilting function, characterizes the how that original external population is tilted to generate the new target population. The general class of population weighted average treatment effect (PWATE) is given by
\begin{equation}\label{balancingweights}
    \tau_{h}=\frac{E\left[h(\mathbf{X}) \tau(\mathbf{X}) \right]}{E\left[h(\mathbf{X})\right]}.
\end{equation}
In this equation, the expectations in both the numerator and the denominator are taken over the target population. The tilting function \(h(\mathbf{X})\) is a non-negative function of covariates and satisfies \(E\left[h(\mathbf{X})\right] > 0\). When $h(\mathbf{X})=1$ and $h(\mathbf{X})\propto e_{\text{sp}}(\mathbf{X})$, the PWATE estimand $\tau_{h}$ reduces to PATE and PATT, respectively.

\subsection{Incorporating survey weights into propensity score weighting with balancing weights}\label{sec2.2}

Under the balancing weights framework, the tilting function $h(\mathbf{X}) = g(\mathbf{X})/f(\mathbf{X})$ is used to tilt the covariate distribution in the population $f(\mathbf{X})$ to that in the target population $g(\mathbf{X})$. 
Since our goal is to derive the expressions of various estimands for target populations in the original population, we first provide a general identification formula of this class of estimands $\tau_{h}(\mathbf{\mathbf{X}})$ considering the average treatment effect over the target population in the original population. Under the retrospective design, Equation \eqref{balancingweights} can be expressed as, 
\begin{equation}
\begin{aligned}
\tau_h
&= \frac{E\left[\displaystyle\frac{h(\mathbf{X})Z  Y}{p_{1}(\mathbf{X})  e_{\text{sp}}(\mathbf{X})}   \Bigg | S=1 \right]}{E\left[\displaystyle\frac{h(\mathbf{X})Z}{p_{1}(\mathbf{X})  e_{\text{sp}}(\mathbf{X})}  \Bigg | S=1 \right]} - \frac{E\left[\displaystyle\frac{h(\mathbf{X})(1-Z)  Y}{p_{0}(\mathbf{X})  \left(1 - e_{\text{sp}}(\mathbf{X})\right)}   \Bigg | S=1 \right]}{E\left[\displaystyle\frac{h(\mathbf{X})(1-Z)}{p_{0}(\mathbf{X})  \left(1 - e_{\text{sp}}(\mathbf{X})\right)}   \Bigg| S=1 \right]}.
\end{aligned}
\end{equation}
When the expectations are replaced with sample means, an unbiased estimator of $\tau_{h}$ in the form of a weighted difference is expressed as 
\begin{equation}
\begin{aligned} \label{tauhPointEstimator}
\widehat{\tau}_{h} &= \frac{\sum_{i=1}^{n}\omega_{1,p}(\mathbf{X}_i) Z_{i} Y_{i} }{\sum_{i=1}^{n}\omega_{1,p}(\mathbf{X}_i) Z_{i} } - \frac{\sum_{i=1}^{n} \omega_{0,p}(\mathbf{X}_i)(1-Z_{i}) Y_{i} }{\sum_{i=1}^{n}  \omega_{0,p}(\mathbf{X}_i)(1-Z_{i})}.
\end{aligned}
\end{equation}
with weights given by 
\begin{equation}
\begin{aligned} \label{balancingWeightsPair}
(\omega_{1,p}(\mathbf{X}_i), \omega_{0,p}(\mathbf{X}_i)) = \left(\displaystyle\frac{h(\mathbf{X}_i)}{p_{1}(\mathbf{X}_i)  e_{\text{sp}}(\mathbf{X}_i)}, \displaystyle\frac{h(\mathbf{X}_i)}{p_{0}(\mathbf{X}_i)  \left(1 - e_{\text{sp}}(\mathbf{X}_i)\right)}\right).
\end{aligned}
\end{equation}
The derivation is included in Web Appendix A.1. Although this expression is specialized to retrospective designs, it can be modified to address prospective designs by defining the sampling weights as \(p(S_i=1 \mid \mathbf{X}_i)\) for individual $i$.  
In such cases, one can factorize \(p(Z_i=1, S_i=1 \mid \mathbf{X}_i)\) differently as \(p(S_i=1 \mid \mathbf{X}_i)  p(Z_i=1 \mid S_i=1, \mathbf{X}_i)\), where the second term represents the sample-level propensity score.\cite{salerno2025whatsweight} Further details on the corresponding estimators under prospective designs are discussed in Web Appendix A.6.

The specification of \(h(\mathbf{X})\) can adopt various forms depending on specific statistical or scientific considerations. Some commonly used forms of balancing weights and their associated target populations, tilting functions, and estimands are presented in Table \ref{Table1CommonBalancingWeightsUnderSurveySettings}. 
For example, we obtain PATE defined as \(\tau_{\text{PATE}} = E[Y(1) - Y(0)]\), when \(g(\mathbf{X}) = f(\mathbf{X})\) and \(h(\mathbf{X}) = 1\). We obtain PATT expressed as \(\tau_{\text{PATT}} = E[Y(1) - Y(0) \mid Z = 1]\), when \(g(\mathbf{X}) = f(\mathbf{X} \mid Z = 1)\) and \(h(\mathbf{X}) = e_{\text{sp}}(\mathbf{X})\). Complementary to PATT, PATC represents the Population Average Treatment Effect for the Control, given by \(\tau_{\text{PATC}} = E[Y(1) - Y(0) \mid Z = 0]\), with \(g(\mathbf{X}) = f(\mathbf{X} \mid Z = 0)\) and \(h(\mathbf{X}) = 1 - e_{\text{sp}}(\mathbf{X})\). 
Importantly, these different tilting functions target subpopulations within the original external population from which the survey sample is drawn and hence the actual target population can differ from that original survey population. In our context, the survey weights are used to inform the original external population from which the survey sample is drawn, whereas $h(\mathbf{X})$ determines which subpopulation receives emphasis in the causal effect estimation. Even when the external population is well defined through a probability sampling design, causal identification from observational data still hinges on adequate overlap in treatment assignment within that population. The overlap and truncated combined estimands below formalize this principle.
Specifically, for the truncated combined approach, the tilting function \(h(\mathbf{X}) = 1\{\alpha <e_{\text{sp}}(\mathbf{X}) < 1 - \alpha\}\) can be used to exclude individuals with extreme propensity scores and to target a population with sufficient overlap.\cite{Crump2009, Traskin2011} This defines a ``region of common support'' within the original external population where treatment effects are identifiable without extrapolation, while preventing the multiplicative impact due to extreme survey weights $1/p_z(\mathbf{X})$ and extreme propensity score weights $1/e_{sp}(\mathbf{X})$ or $1/(1-e_{sp}(\mathbf{X}))$ that would otherwise result in bias and inflate variance. The parameter \(\alpha\), typically chosen within the range \((0, 0.1)\), indicates the truncation threshold for propensity scores. 

The overlap weights defined by Li et al. \cite{Li2018} correspond to the Average Treatment Effect among the Overlap Population (PATO) under the tilting function of the average \(h(\mathbf{X}) = e_{\text{sp}}(\mathbf{X}) \left(1-e_{\text{sp}}(\mathbf{X})\right)\), such that greater emphasis is given to units with propensity scores close to 0.5 and smaller weights are assigned to those with extreme propensity scores with more deterministic treatment assignment. In survey observational studies, this overlap population is defined relative to the survey's original external population and still includes individuals whose treatment decisions are genuinely uncertain---the equipoise region where both treatments are plausible options and treatment effect evidence is most valuable for informing policy or clinical decisions.
With additional regularity assumptions, we show that the original minimum asymptotic variance property of overlap weights continues to hold, along with the exact balance property under survey-weighted logistic propensity score modeling. We summarize the properties below and the proof can be found in Web Appendix A.2.

\begin{table*}[htbp]
\scriptsize 
\setlength{\tabcolsep}{3pt} 
\renewcommand{\arraystretch}{1.2} 
\caption{A summary of different forms of balancing weights, target population, tilting function, and estimand for survey observational studies.}\label{Table1CommonBalancingWeightsUnderSurveySettings}
\[
\renewcommand{\arraystretch}{1.2} 
\setlength{\tabcolsep}{0.8pt} 
\begin{array}{lccc}
\hline \rowcolor[rgb]{0.92,0.92,0.92} \text {Target population} & h(\mathbf{X}) & \text { Estimand } & \text { Weight }\left(\omega_{1,p}, \omega_{0,p}\right) \\
\hline 
\rule{0pt}{12pt} \begin{array}{l}
\text {Combined }
\end{array} & 1  & \text { PATE }  & \left(\displaystyle \frac{1}{p_1(\mathbf{X})}  \displaystyle \frac{1}{e_{\text{sp}}\left(\mathbf{X}\right)}, \displaystyle \frac{1}{p_0(\mathbf{X})}  \displaystyle \frac{1}{1-e_{\text{sp}}\left(\mathbf{X}\right)}\right) \\
\rowcolor[rgb]{0.92,0.92,0.92}
\rule{0pt}{20pt} \text { Treated } &   e_{\text{sp}}(\mathbf{X}) & \text { PATT } & \left(\displaystyle \frac{1}{p_1(\mathbf{X})}, \displaystyle \frac{1}{p_0(\mathbf{X})}  \frac{e_{\text{sp}}(\mathbf{X})}{1-e_{\text{sp}}(\mathbf{X})}\right) \\
\rule{0pt}{20pt} \text { Control } & (1-e_{\text{sp}}(\mathbf{X})) & \text { PATC } & \left(\displaystyle \frac{1}{p_1(\mathbf{X})} \frac{1-e_{\text{sp}}(\mathbf{X})}{e_{\text{sp}}(\mathbf{X})}, \displaystyle \frac{1}{p_0(\mathbf{X})}\right) \\
\rowcolor[rgb]{0.92,0.92,0.92}
\rule{0pt}{20pt} \begin{array}{c}
\text {Trimmed } \\
\text {combined }
\end{array} & \mathbf{1}(\alpha<e_{\text{sp}}(\mathbf{X})<1-\alpha) & &\left(\displaystyle \frac{1}{p_1(\mathbf{X})} \frac{1(\alpha<e_{\text{sp}}(\mathbf{X})<1-\alpha)}{e_{\text{sp}}(\mathbf{X})}, \displaystyle \frac{1}{p_{0}(\mathbf{X})}  \frac{1(\alpha<e_{\text{sp}}(\mathbf{X})<1-\alpha)}{1-e_{\text{sp}}(\mathbf{X})}\right) \\
\rule{0pt}{20pt} \text { Overlap } &  e_{\text{sp}}(\mathbf{X})   \left(1-e_{\text{sp}}(\mathbf{X})\right) & \text { PATO } & \left(\displaystyle \frac{1}{p_1(\mathbf{X})}   \left(1-e_{\text{sp}}(\mathbf{X})\right), \displaystyle \frac{1}{p_{0}(\mathbf{X})}  e_{\text{sp}}(\mathbf{X})\right) \\
\hline
\end{array}
\]
\end{table*}

\begin{proposition}\label{ExactBalance}
\emph{When the propensity scores are estimated using a survey-weighted logistic regression under weighted pseudo maximum likelihood estimation, where \(\widehat{e}_{\text{sp}}(\mathbf{X}_i) = \operatorname{logit}^{-1}(\widehat{\boldsymbol{\beta}}_0 + \mathbf{X}_i \widehat{\boldsymbol{\beta}}_{sw}^\top)\), the application of overlap weights ensures exact mean balance of any given covariate between groups in the target population. That is,
\begin{align*}
\frac{\sum_{i=1}^{n} X_{ik} Z_i (1-\widehat{e}_{\text{sp}}(\mathbf{X}_i))/p_1(\mathbf{X}_i)}{\sum_{i=1}^{n} Z_i (1-\widehat{e}_{\text{sp}}(\mathbf{X}_i))/p_1(\mathbf{X}_i)} = \frac{\sum_{i=1}^{n} X_{ik} (1-Z_i)  \widehat{e}_{\text{sp}}(\mathbf{X}_i)/p_0(\mathbf{X}_i)}{\sum_{i=1}^{n} (1-Z_i) \widehat{e}_{\text{sp}}(\mathbf{X}_i)/p_0(\mathbf{X}_i)}, \quad \text{for} \quad k=1,2, \ldots, K. 
\end{align*}}
\end{proposition}

In the absence of survey sampling or under simple random sampling, overlap weights also possess the optimal efficiency property, as proved in Li et al.\cite{Li2018} Under a retrospective survey sampling design, in Web Appendix A.2 (Proposition S1), we prove that when the standard homoskedasticity assumption holds, that is, $\text{Var}(Y(z) \mid \mathbf{X}) = \sigma^2$, the theoretically variance-minimizing tilting function is in fact $h^*(\mathbf{X}) \propto \left(p_1(\mathbf{X})p_0(\mathbf{X})/p(\mathbf{X})\right) e_{\text{sp}}(\mathbf{X})(1-e_{\text{sp}}(\mathbf{X}))$, which depends on the treatment-specific sampling probabilities $p_1(\mathbf{X})$ and $p_0(\mathbf{X})$. While this tilting function is theoretically appealing, it is less interpretable in practical applications. On the other hand, the original tilting function that generates the standard overlap weights, $h(\mathbf{X}) \propto e_{\text{sp}}(\mathbf{X})(1-e_{\text{sp}}(\mathbf{X}))$, attains the minimum variance property under two conditions: (i) simple random sampling where $p_1(\mathbf{X}) = p_0(\mathbf{X}) = p$ is constant, with the standard homoskedasticity assumption, or (ii) covariate-dependent sampling with a modified homoskedasticity condition $\text{Var}(Y(z)/\sqrt{p_z(\mathbf{X})} \mid \mathbf{X}) = \tilde{\sigma}^2$. Due to interpretability, we focus on the standard overlap weights throughout. While the modified homoskedasticity condition in (ii) is primarily for theoretical purposes and may not hold in practice, our simulations in Section \ref{sec4} demonstrate that overlap weights can still achieve smaller variances than inverse probability weights in finite samples.

\subsection{Asymptotic normality and the sandwich variance estimator}\label{sec2.3}

One viable approach for statistical inference is the bootstrap method, which could ascertain the variance of \(\widehat{\tau}_{h}\) across various target populations. However, bootstraps can involve a substantial amount of computation, especially when the sample size gets larger. In addition, bootstrap survey data often entails additional complications, see, for example, Ackerman et al.\cite{Ackerman2021} We provide a closed-form variance estimator via the empirical sandwich method, which serves as a feasible solution to circumvent the need for extensive computation.\cite{stefanski2002calculus} Detailed derivations can be found in Web Appendix A.3.

Assuming the propensity scores are estimated using a survey-weighted logistic regression model and in the form of \( e_{\text{sp}}\left(\mathbf{X}_{i}; \boldsymbol{\beta}\right) = 1/\left(1+\exp\left(-\mathbf{X}_{i}^{\top}\boldsymbol{\beta}\right)\right)\). We define \(\tau_{1}={E\left[h(\mathbf{X}) \mu_{1}(\mathbf{X}) \right]}/{E\left[h(\mathbf{X}) \right]}\) and \(\tau_{0}={E\left[h(\mathbf{X}) \mu_{0}(\mathbf{X}) \right]}/{E\left[h(\mathbf{X}) \right]}\), where  \(\mu_{1}(\mathbf{X}) = E\left[Y(1) \mid \mathbf{X}\right]\) and \(\mu_{0}(\mathbf{X}) = E\left[Y(0) \mid \mathbf{X}\right]\), as the treatment-specific average potential outcomes in the target population, and the treatment effect estimator of target population can be written as \(\widehat{\tau}_{h} = \widehat{\tau}_1 - \widehat{\tau}_0\). 
The consistent sandwich-type variance estimator for \(\widehat{\tau}_h\) is given by:
\begin{equation}
\begin{aligned} \label{SandwichVarianceEstimator}
\widehat{\text{Var}}(\widehat{\tau}_h) = n^{-2} \sum_{i=1}^n \left( \widehat{I}_i / \widehat{\nu} \right)^2,
\end{aligned}
\end{equation}
where \(\widehat{\nu} = n^{-1} \sum_{i=1}^n h(\mathbf{X}_i)/p(\mathbf{X}_i)\). Here, the influence function \(\widehat{I}_i\) characterizing the contribution of the \(i\)-th sampled individual to the effect estimator is defined as:
\begin{equation}
\begin{aligned}
\widehat{I}_i = Z_i \widehat{\omega}_{1,p}(\mathbf{X}_i; \widehat{\boldsymbol{\beta}}) \left(Y_i - \widehat{\tau}_1\right) - (1 - Z_i) \widehat{\omega}_{0,p}(\mathbf{X}_i; \widehat{\boldsymbol{\beta}}) \left(Y_i - \widehat{\tau}_0\right)+ \widehat{\mathbf{H}}_{\widehat{\boldsymbol{\beta}}}^{\top} \widehat{E}_{\widehat{\boldsymbol{\beta}} \widehat{\boldsymbol{\beta}}}^{-1} \frac{1}{p_z(\mathbf{X}_i)} \left(Z_i - \widehat{e}_{\text{sp}}(\mathbf{X}_i; \widehat{\boldsymbol{\beta}})\right) \mathbf{X}_i.
\end{aligned}
\end{equation}
where the gradient of the balancing weights with respect to the logistic regression parameters \(\boldsymbol{\beta}\) is:
\begin{equation}
\begin{aligned}
\widehat{\mathbf{H}}_{\widehat{\boldsymbol{\beta}}} = n^{-1} \sum_{i=1}^n \left[Z_i \left(Y_i - \widehat{\tau}_1\right) \omega_{1,p,\boldsymbol{\beta}}^{\prime} - (1 - Z_i) \left(Y_i - \widehat{\tau}_0\right) \omega_{0,p,\boldsymbol{\beta}}^{\prime}\right],
\end{aligned}
\end{equation}
Here, \(\omega_{1,p,\boldsymbol{\beta}}^{\prime} = \partial \omega_{1,p}(\mathbf{X}_{i}; \boldsymbol{\beta})/ \partial \boldsymbol{\beta}\) and \(\omega_{0,p,\boldsymbol{\beta}}^{\prime} =\partial \omega_{0,p}(\mathbf{X}_{i}; \boldsymbol{\beta})/ \partial \boldsymbol{\beta}\) quantify how sensitive the balancing weights are to changes in propensity score model parameters. Further, the covariance matrix of the propensity score estimates \(\widehat{\mathbf{E}}_{\boldsymbol{\beta} \boldsymbol{\beta}}\) here is expressed as:
\begin{equation}
\begin{aligned}
\widehat{E}_{\widehat{\boldsymbol{\beta}} \widehat{\boldsymbol{\beta}}} = n^{-1} \sum_{i=1}^n \frac{1}{p_z(\mathbf{X}_i)} \widehat{e}_{\text{sp}}(\mathbf{X}_i; \widehat{\boldsymbol{\beta}}) \left(1 - \widehat{e}_{\text{sp}}(\mathbf{X}_i; \widehat{\boldsymbol{\beta}})\right) \mathbf{X}_i \mathbf{X}_i^{\top}.
\end{aligned}
\end{equation}
This sandwich variance estimator extends the one in Li et al.\cite{li2019addressing} by accounting for the known survey weights and the uncertainty in the estimation of survey-weighted propensity scores.

\section{Improving the weighting estimator through outcome regression}\label{sec3}

In this section, we first develop the augmented weighting (AW) estimating equations based on the balancing weights in Section \ref{sec3.1} in survey observational data. We then introduce three versions of augmented estimators in Sections \ref{sec3.2} to \ref{sec3.4}: the one-step moment estimator, the clever covariate estimator, and the weighted regression estimator. Each augmented weighting method is discussed with a focus on how survey weights are integrated at various stages. Section \ref{sec3.5} presents the closed-form sandwich variance estimation for the augmented estimators for computationally efficient inference. This section primarily focuses on the development of estimators under the retrospective case, consistent with our previous notation, while modifications required for prospective designs are outlined in the Web Appendix A.6.

\subsection{Augmented weighting estimating equations for survey observational data}\label{sec3.1}

To improve efficiency over weighting alone, we extend the asymptotic properties of augmented estimators\cite{mao2019propensity} following §13.5 of Tsiatis \cite{Tsiatis2006SemiparametricTA} to formally incorporate survey weights under the retrospective design. 
Throughout this section, $\tau_h$ denotes the target estimand in \eqref{balancingweights}, $\widehat{\tau}_h^{\text{AW}}$ denotes the augmented weighting estimator, and $\phi_{h,i}^{\text{AW}}$ denotes the influence function contribution from subject $i$, so that $\sqrt{n}\{\widehat{\tau}_h^{\text{AW}}-\tau_h\}=n^{-1/2}\sum_{i=1}^n \phi_{h,i}^{\text{AW}}+o_p(1)$.
In Web Appendix A.4
, we first establish the following result. 

\begin{proposition}\label{RALProp}
\emph{If the survey weights $1/p_z(\mathbf{X})$ and the propensity scores $e_{\text{sp}}(\mathbf{X})$ are known, the class of observed sample data influence functions for regular asymptotically linear estimators of population estimand $\tau_h$ is,
\begin{align} \label{eq:IF-class}
\phi_h^{\text{AW}} &= \frac{h(\mathbf{X})/p(\mathbf{X})}{\Psi}\left[
\frac{Z Y p(\mathbf{X})}{p_1(\mathbf{X}) e_{\text{sp}}(\mathbf{X})} 
- \frac{(1-Z) Y p(\mathbf{X})}{p_0(\mathbf{X}) (1 - e_{\text{sp}}(\mathbf{X}))} 
\right] - \tau_h + \left\{Z - \frac{p_1(\mathbf{X}) e_{\text{sp}}(\mathbf{X})}{p(\mathbf{X})}\right\} \varphi(\mathbf{X}),
\end{align}
for any function $\varphi(\mathbf{X})$, where $\Psi = E[h(\mathbf{X})/p(\mathbf{X}) \mid S=1]$.}
\end{proposition}

Here $\Psi$ is the normalization term and the tilting function $h(\mathbf{X})$
can depend on the population propensity score \(e_{\text{sp}}(\mathbf{X})\), as illustrated in Table \ref{Table1CommonBalancingWeightsUnderSurveySettings}.
The augmentation term $\{Z - p_1(\mathbf{X})e_{\text{sp}}(\mathbf{X})/p(\mathbf{X})\}\varphi(\mathbf{X})$ has mean zero and captures the additional efficiency gain from outcome modeling. 
Based on this proposition, one can follow the argument in Tsiatis \cite{Tsiatis2006SemiparametricTA} and show that, among all estimators whose influence functions fall within this class, the augmented estimator with  
\begin{align}
\varphi(\mathbf{X}) 
& = -\frac{h(\mathbf{X}) / p(\mathbf{X})}{\Psi}
\left[ 
\frac{p(\mathbf{X}) E[Y \mid Z = 1, \mathbf{X}, S = 1]}{p_1(\mathbf{X}) e_{\text{sp}}(\mathbf{X})}
+ 
\frac{p(\mathbf{X}) E[Y \mid Z = 0, \mathbf{X}, S = 1]}{p_0(\mathbf{X}) (1 - e_{\text{sp}}(\mathbf{X}))}
\right],
\end{align}
achieves the smallest asymptotic variance, where $E[Y \mid Z=z, \mathbf{X}, S=1]$ is modeled by correctly specified outcome models for $z = 1$ and $0$.
This motivates the augmented weighting estimating equation tailored to survey observational data, where the estimator $\widehat{\tau}_h^{\text{AW}}$ solves:


\begin{align}\label{AWestimatingequationSection3.1}
\frac{1}{\Psi}  \frac{1}{n} \sum_{i=1}^{n} \Bigg\{ & \frac{h(\mathbf{X}_i)}{p(\mathbf{X}_i)}  (m_1(\mathbf{X}_i) - m_0(\mathbf{X}_i)) + \frac{Z_i  h(\mathbf{X}_i)}{e_{\text{sp}}(\mathbf{X}_i)  p_{1}(\mathbf{X}_i)}  (Y_i - m_1(\mathbf{X}_i)) \nonumber \\
& - \frac{(1 - Z_i)  h(\mathbf{X}_i)}{(1 - e_{\text{sp}}(\mathbf{X}_i))  p_{0}(\mathbf{X}_i)}  (Y_i - m_0(\mathbf{X}_i)) - \tau_h  \Bigg\} = 0,
\end{align}

where $m_z(\mathbf{X}) = E[Y \mid \mathbf{X}, Z=z, S=1]$ represents the outcome regression for $z = 1$ and $0$.
By substituting $(\omega_{1,p}(\mathbf{X}), \omega_{0,p}(\mathbf{X}))$ defined in Equation \eqref{balancingWeightsPair} and $\Psi = E[h(\mathbf{X})/p(\mathbf{X}) \mid S=1] = E[Z \omega_{1,p}(\mathbf{X}) \mid S=1] = E[(1-Z) \omega_{0,p}(\mathbf{X}) \mid S=1]$ into above Equation \eqref{AWestimatingequationSection3.1}, 
the solution to this estimating equation is the augmented weighting estimator,

\begin{align}\label{AWGeneralEstimatorSW}
\widehat{\tau}_{h}^{\text{AW}} &= \frac{\sum_{i=1}^{n} h(\mathbf{X}_i)/p(\mathbf{X}_i) \left(\widehat{m}_1(\mathbf{X}_i) - \widehat{m}_0(\mathbf{X}_i)\right)}{\sum_{i=1}^{n} h(\mathbf{X}_i)/p(\mathbf{X}_i)} \\\nonumber
&\quad + \frac{\sum_{i=1}^{n} Z_i  \omega_{1,p}(\mathbf{X}_i)  \left(Y_i - \widehat{m}_1(\mathbf{X}_i)\right)}{\sum_{i=1}^{n} Z_i  \omega_{1,p}(\mathbf{X}_i)} - \frac{\sum_{i=1}^{n} (1-Z_i)  \omega_{0,p}(\mathbf{X}_i)  \left(Y_i - \widehat{m}_0(\mathbf{X}_i)\right)}{\sum_{i=1}^{n} (1-Z_i)  \omega_{0,p}(\mathbf{X}_i)}.
\end{align}
The first term represents the outcome-model-based prediction of $\tau_h$, 
while the second and third terms provide bias correction through weighted 
residuals when the outcome models are misspecified.
Under retrospective designs, when the marginalized survey weights \(1/p(\mathbf{X}_i)\) are not directly accessible, they can be approximated by \(p(\mathbf{X}_i) = r_z(\mathbf{X}_i) p_z(\mathbf{X}_i)\) for each individual with \(z_i = 1\) or \(0\), as defined in Section \ref{sec2.1}.
It is worth noting that estimating $r_z(\mathbf{X})$ requires specification of not only the population-level propensity score $e_{\text{sp}}(\mathbf{X})$ but also the sample-level propensity score $e_{\text{fp}}(\mathbf{X})$, whose form becomes a nonlinear logistic under retrospective treatment-dependent sampling when the true population-level treatment assignment follows a standard logistic model, as shown in Equation \eqref{eq:efp-retro-ps}. In Web Appendix B.3, we propose two strategies to resolve this issue: (i) a practical approximation using standard unweighted logistic regression, which can be justified under low sampling rates (generally below 10\%) typical of large-scale surveys where log-linear and logistic models are approximately equivalent, and (ii) a nonlinear maximum likelihood approach using the induced form derived from Equation \eqref{eq:efp-retro-ps}, which requires explicit modeling of the sampling mechanism and may be preferred when the sampling rate is much higher.

\begin{proposition}\label{ConsistencyProp}
\emph{The augmented weighting estimator $\widehat{\tau}_{h}^{\text{AW}}$ remains consistent for $\tau_h$ provided that the population-level propensity score model is correctly specified, irrespective of the accuracy of the outcome models. When the population-level propensity score model $e_{\text{sp}}(\mathbf{X})$ is misspecified but the outcome models are correct, this estimator consistently estimates $\widetilde{\tau}_h$.}
\end{proposition}

Here $\widetilde{\tau}_{h}= {E\left[h\left(e_{\text{sp}}(\mathbf{X}; \widetilde{\boldsymbol{\beta}})\right) \tau(\mathbf{X})\right]}/{E\left[h(e_{\text{sp}}(\mathbf{X}; \widetilde{\boldsymbol{\beta}}))\right]}$ represents the average treatment effect in a shifted target population determined by the misspecified propensity score model $e_{\text{sp}}(\mathbf{X}; \widetilde{\boldsymbol{\beta}})$. For inverse probability weighting (IPW) where $h(\mathbf{X}) = 1$, $\widetilde{\tau}_h$ reduces to $\tau_h$ and the estimator achieves double robustness. In contrast, when $h(\mathbf{X})$ depends on $e_{\text{sp}}(\mathbf{X})$ (e.g., PATT, PATO), the estimator is singly robust with respect to the propensity score model. The proof is provided in Web Appendix A.4.
Although the shifted estimand $\widetilde{\tau}_h$ targets a different (and generally not interpretable) population when the propensity score is misspecified, Zhou et al.\cite{Zhou2020SMMR} empirically demonstrated that overlap weights exhibit smaller bias than IPW under propensity score model misspecification. Our simulation studies in Section \ref{sec4} demonstrate that this observation continues to hold in survey sampling settings, with augmented estimators maintaining robust performance even when propensity score models are misspecified.

In the following, we introduce three variants of the augmented weighting estimators that are applicable to survey observational data, all connected to Equation \eqref{AWGeneralEstimatorSW}.

\subsection{One-step moment estimator}\label{sec3.2}

The first augmented estimator is the one-step moment estimator (MOM). Under the balancing weights framework, the moment estimator can be adapted for any specified target population defined by \( h(\mathbf{X}) \).\cite{mao2019propensity} We denote the regression model as the conditional expectation of the outcome given covariates over treatment and control group in the observed sample as \(m_{1}^{\text{MOM}}(\mathbf{X}; \mathbf{\alpha}_1)\) and \(m_{0}^{\text{MOM}}(\mathbf{X}; \mathbf{\alpha}_0)\), with \(\mathbf{\alpha}_1\) and \(\mathbf{\alpha}_0\) as vectors of regression parameters. By plugging in \(m_{1}^{\text{MOM}}(\mathbf{X}; \widehat{\mathbf{\alpha}}_1)\) and \(m_{0}^{\text{MOM}}(\mathbf{X}; \widehat{\mathbf{\alpha}}_0)\) into \(\widehat{\tau}_{h}^{\text{AW}}\), the moment estimator can be expressed as,
\begin{align}\label{tau_mom}
\widehat{\tau}_{h}^{\text{MOM}}  &= \frac{\sum_{i=1}^n h(\mathbf{X}_i)/ p(\mathbf{X}_i)\left[m_{1}^{\text{MOM}}(\mathbf{X}_i; \widehat{\mathbf{\alpha}}_1)-m_{0}^{\text{MOM}}(\mathbf{X}_i; \widehat{\mathbf{\alpha}}_0)\right]}{\sum_{i=1}^n  h(\mathbf{X}_i)/p(\mathbf{X}_i)} \\\nonumber
&\quad +\frac{\sum_{i=1}^n \omega_{1,p}(\mathbf{X}_i) Z_i\left[Y_i-m_{1}^{\text{MOM}}(\mathbf{X}_i; \widehat{\mathbf{\alpha}}_1)\right]}{\sum_{i=1}^n \omega_{1,p}(\mathbf{X}_i) Z_i} - \frac{\sum_{i=1}^n \omega_{0,p}(\mathbf{X}_i)(1-Z_i)\left[Y_i-m_{0}^{\text{MOM}}(\mathbf{X}_i; \widehat{\mathbf{\alpha}}_0)\right]}{\sum_{i=1}^n \omega_{0,p}(\mathbf{X}_i)(1-Z_i)}.
\end{align}

In this estimator, survey weights must be incorporated in three places. First, survey weights are used to estimate the population-level propensity score model $e_{\text{sp}}(\mathbf{X}; \boldsymbol{\beta})$, via weighted logistic regression. Second, the marginalized sampling probability $p(\mathbf{X}_i)$ appears in the first term of Equation \eqref{tau_mom} through the ratio $h(\mathbf{X}_i)/p(\mathbf{X}_i)$, which reweights the outcome model predictions to the target population.
Third, the treatment-specific sampling probabilities $p_z(\mathbf{X}_i)$ are incorporated into the balancing weight pair, $\omega_{1,p}(\mathbf{X}_i) = h(\widehat{e}_{\text{sp}}(\mathbf{X}_i; \widehat{\boldsymbol{\beta}}))/(p_{1}(\mathbf{X}_i)  \widehat{e}_{\text{sp}}(\mathbf{X}_i; \widehat{\boldsymbol{\beta}}))$ and $\omega_{0,p}(\mathbf{X}_i) = h(\widehat{e}_{\text{sp}}(\mathbf{X}_i; \widehat{\boldsymbol{\beta}}))/(p_{0}(\mathbf{X}_i) (1 - \widehat{e}_{\text{sp}}(\mathbf{X}_i; \widehat{\boldsymbol{\beta}})))$, which in the second and third terms of Equation \eqref{tau_mom} standardize the weighted residuals for bias correction.
Here, the conditional expectations, \( m_{1}^{\text{MOM}}(\mathbf{X}; \mathbf{\alpha}_1) \) and \( m_{0}^{\text{MOM}}(\mathbf{X}; \mathbf{\alpha}_0) \), are regression models to recover the missing potential outcomes, and these models are fitted to the observed data without requiring adjustment for the survey sampling design. Detailed proof for the consistency of \(\widehat{\tau}_{h}^{\text{MOM}} \) is in Web Appendix A.5.

Upon closer inspection, Equation \eqref{tau_mom} can be decomposed into two components: \(\widehat{\tau}_{h}^{\text{MOM}}  = \widehat{\mu}_{1}^{\text{MOM}} - \widehat{\mu}_{0}^{\text{MOM}}\), where 
\begin{align}
\widehat{\mu}_{1}^{\text{MOM}}  &= \frac{\sum_{i=1}^n h(\mathbf{X}_i)/ p(\mathbf{X}_i)m_{1}^{\text{MOM}}(\mathbf{X}_i; \widehat{\mathbf{\alpha}}_1)}{\sum_{i=1}^n  h(\mathbf{X}_i)/p(\mathbf{X}_i)}  +\frac{\sum_{i=1}^n \omega_{1,p}(\mathbf{X}_i) Z_i\left[Y_i-m_{1}^{\text{MOM}}(\mathbf{X}_i; \widehat{\mathbf{\alpha}}_1)\right]}{\sum_{i=1}^n \omega_{1,p}(\mathbf{X}_i) Z_i},
\end{align}
and $\widehat{\mu}_{0}^{\text{MOM}}$ is the counterpart for the control group.
Each of $\widehat{\mu}_{1}^{\text{MOM}}$ and $\widehat{\mu}_{0}^{\text{MOM}}$ is characterized by a regression model enhanced by weighted residuals for each group. Regarding the augmentation properties of \(\widehat{\tau}_{h}^{\text{MOM}} \), consider the expression for \(\widehat{\mu}_{1}^{\text{MOM}}\), which converges in probability to
\begin{align} \label{MOMDoublyRobust}
\widehat{\mu}_{1}^{\text{MOM}} 
\xrightarrow{p} \frac{E\left[h(\mathbf{X}) Y(1)\right]}{E\left[h(\mathbf{X})\right]} +\frac{1}{E\left[h(\mathbf{X})\right]} E\left[h(\mathbf{X})\left(\frac{Z}{e_{\text{sp}}(\mathbf{X})}-1\right)\left(Y(1)-m_1^{\text{MOM}}(\mathbf{X};\mathbf{\alpha}_1)\right)\right].
\end{align}
The second term in this equation vanishes when either the postulated regression model \(m_{1}^{\text{MOM}}(\mathbf{X}; \mathbf{\alpha}_1)\) or the survey-weighted propensity score model \(e_{\text{sp}}(\mathbf{X}; \boldsymbol{\beta})\) is correctly specified. 
When the tilting function \(h(\mathbf{X})\) is specified as a constant proportional function, such as \(h(\mathbf{X}) \propto 1\), it corresponds to IPW and defines the entire population as the target population (i.e., PATE is the estimand). Under this configuration, \(\widehat{\tau}_{h}^{\text{MOM}}\) exhibits double robustness, meaning that consistency is guaranteed if either the outcome model \(m_{1}^{\text{MOM}}(\mathbf{X}; \mathbf{\alpha}_1)\) or the propensity score model \(e_{\text{sp}}(\mathbf{X}; \boldsymbol{\beta})\) is correctly specified.

In more general cases, when the tilting function \(h(\mathbf{X})\) explicitly depends on \(e_{\text{sp}}(\mathbf{X})\), as illustrated in Table \ref{Table1CommonBalancingWeightsUnderSurveySettings}, the estimator \(\widehat{\mu}_{1}^{\text{MOM}}\) is only singly robust. In this context, consistency is ensured only if the propensity score model \(e_{\text{sp}}(\mathbf{X}; \boldsymbol{\beta})\) is correctly specified, which allows for unbiased estimation of the first term in Equation \eqref{MOMDoublyRobust}. However, as demonstrated by Mao et al.,\cite{mao2019propensity} integrating an outcome model into the weighting estimator can improve efficiency compared to weighting alone. Through simulation studies, we demonstrate in Section \ref{sec4} that this observation continues to hold in survey observational studies.

\subsection{Clever covariate regression estimator}\label{sec3.3}

An alternative strategy for constructing an augmented estimator involves incorporating functions of estimated propensity scores into the outcome regression model as an additional covariate. Scharfstein et al.\cite{scharfstein1999adjusting} introduced a specific version of this approach by adding the inverse propensity score into outcome models. Later, Bang et al.\cite{BangRobins2005} demonstrated that this estimator is essentially a solution to the augmented weighting estimating equation; hence in the context of IPW, the resulting estimator is also doubly robust. This method is also referred to as the ``clever covariate'' estimator (CVR), and we present a version of this estimator for survey observational studies. 
Specifically, we first fit the following modified outcome models \(m_1(\mathbf{X})\) and \(m_0(\mathbf{X})\) with the additional clever covariate:
\begin{align*}
m_{1}^{\text{CVR}}\left(\mathbf{X}; \widehat{\boldsymbol{\alpha}}_{1}^{\text{CVR}}\right) &= g^{-1}\{m_1(\mathbf{X}; \widehat{\boldsymbol{\alpha}}_1) + \widehat{\phi}_1\omega_{1,p}(\mathbf{X})\} = g^{-1}\left\{m_1(\mathbf{X}; \widehat{\boldsymbol{\alpha}}_1) + \widehat{\phi}_1 h(\mathbf{X})/\left(p_{1}(\mathbf{X}) e_{\text{sp}}(\mathbf{X}; \widehat{\boldsymbol{\beta}}) \right)\right\}, \\
m_{0}^{\text{CVR}}\left(\mathbf{X}; \widehat{\boldsymbol{\alpha}}_{0}^{\text{CVR}}\right) &= g^{-1}\{m_0(\mathbf{X}; \widehat{\boldsymbol{\alpha}}_0) + \widehat{\phi}_0\omega_{0,p}(\mathbf{X})\} = g^{-1}\left\{m_0(\mathbf{X}; \widehat{\boldsymbol{\alpha}}_0) + \widehat{\phi}_0 h(\mathbf{X})/\left(p_{0}(\mathbf{X}) (1 - e_{\text{sp}}(\mathbf{X}; \widehat{\boldsymbol{\beta}}))\right)\right\},
\end{align*}
where \(g\) is the canonical link function. The clever covariate estimator is obtained by using the modified regression functions
{\begingroup
\begin{align}
\widehat{\tau}_{h}^{\text{CVR}} & =  \widehat{\tau}_{1}^{\text{CVR}}  - \widehat{\tau}_{0}^{\text{CVR}} \\\nonumber
&= \frac{\sum_{i=1}^{n} \displaystyle h(\mathbf{X}_i)/p(\mathbf{X}_i)  m_{1}^{\text{CVR}}\left(\mathbf{X}_i; \widehat{\boldsymbol{\alpha}}_{1}^{\text{CVR}}\right)}{\sum_{i=1}^{n}  h(\mathbf{X}_i)/p(\mathbf{X}_i)} + \frac{\sum_{i=1}^{n} \omega_{1,p}(\mathbf{X}_i, \widehat{\boldsymbol{\beta}})  Z_i  \left(Y_i - m_{1}^{\text{CVR}}\left(\mathbf{X}_i; \widehat{\boldsymbol{\alpha}}_{1}^{\text{CVR}}\right)\right) }{\sum_{i=1}^{n} \omega_{1,p}\left(\mathbf{X}_i; \widehat{\boldsymbol{\beta}}\right) Z_i} \\\nonumber
& \quad - \frac{\sum_{i=1}^{n} h(\mathbf{X}_i)/p(\mathbf{X}_i) m_{0}^{\text{CVR}}\left(\mathbf{X}_i; \widehat{\boldsymbol{\alpha}}_{0}^{\text{CVR}}\right)}{\sum_{i=1}^{n} h(\mathbf{X}_i)/p(\mathbf{X}_i)}  - \frac{\sum_{i=1}^{n} \omega_{0,p}(\mathbf{X}_i, \widehat{\boldsymbol{\beta}})  (1-Z_i)  \left(Y_i - m_{0}^{\text{CVR}}\left(\mathbf{X}_i; \widehat{\boldsymbol{\alpha}}_{0}^{\text{CVR}}\right)\right)}{\sum_{i=1}^{n} \omega_{0,p}\left(\mathbf{X}_i; \widehat{\boldsymbol{\beta}}\right)(1-Z_i)}. 
\end{align}
\endgroup}The second terms of \(\widehat{\tau}_{1}^{\text{CVR}}\)  and \(\widehat{\tau}_{0}^{\text{CVR}}\) are essentially the score functions of \(m_{1}^{\text{CVR}}\left(\mathbf{X}; \widehat{\boldsymbol{\alpha}}_{1}^{\text{CVR}}\right)\)  and \(m_{0}^{\text{CVR}}\left(\mathbf{X}; \widehat{\boldsymbol{\alpha}}_{0}^{\text{CVR}}\right)\) acting as the weighted residuals of first terms of each, and they become zero when they are correctly specified, simplifying \( \widehat{\tau}_{h}^{\text{CVR}} \) to the following form:
{\begingroup
\begin{align} \label{tauCleverSimplified}
\widehat{\tau}_{h}^{\text{CVR}} = \frac{\sum_{i=1}^{n} \displaystyle h(\mathbf{X}_i)/p(\mathbf{X}_i)  \left(m_{1}^{\text{CVR}}\left(\mathbf{X}_i; \widehat{\boldsymbol{\alpha}}_{1}^{\text{CVR}}\right) -  m_{0}^{\text{CVR}}\left(\mathbf{X}_i; \widehat{\boldsymbol{\alpha}}_{0}^{\text{CVR}}\right)\right)}{\sum_{i=1}^{n}  h(\mathbf{X}_i)/p(\mathbf{X}_i)}. 
\end{align}
\endgroup}
In this clever covariate estimator, survey weights must be incorporated in three places: estimating the propensity score model $e_{\text{sp}}(\mathbf{X}; \boldsymbol{\beta})$, the ratio $h(\mathbf{X}_i)/p(\mathbf{X}_i)$ for reweighting predictions, and the balancing weights $\omega_{z,p}(\mathbf{X}_i)$ serving as clever covariates. The outcome models, including the base models $m_z(\mathbf{X}; \boldsymbol{\alpha}_z)$ and the refitted $m_{z}^{\text{CVR}}\left(\mathbf{X}; \boldsymbol{\alpha}_{z}^{\text{CVR}}\right)$, are fitted without survey weight adjustment.
The clever covariate estimator is asymptotically equivalent to the one-step moment estimator, sharing the same influence function and robustness properties. When $h(\mathbf{X}) \propto 1$, the estimator achieves double robustness. When $h(\mathbf{X})$ depends on $e_{\text{sp}}(\mathbf{X})$, the estimator is singly robust with respect to the propensity score model; if the propensity score model is misspecified, the estimator converges to the shifted estimand $\widetilde{\tau}_{h}$ as discussed in Section \ref{sec3.1}.

In the absence of survey sampling, Kang and Schafer \cite{Kang2007} pointed out that, when the outcome models  \(m_{1}\left(\mathbf{X}; \boldsymbol{\alpha}_{1}\right)\)  and  \(m_{0}\left(\mathbf{X}; \boldsymbol{\alpha}_{0}\right)\) are already correctly specified, the clever covariates as coarse summaries of the covariates may lead to slight overfitting by adding mean-zero noise through weighted residuals. When the survey-weighted propensity score model is correctly specified, the clever covariate method ensures consistent estimation of target parameters, provided that the conditional mean of the outcome varies smoothly with both the survey-weighted propensity score and the sampling probabilities. In theory, this smooth variation remains essential for maintaining the consistency and robustness of the clever covariate method, even in complex survey designs. However, incorporating survey weights twice in constructing the clever covariates may potentially exacerbate the extrapolation problem noted by Robins et al.\cite{Robins2007} because the inflation of extreme value can lead to finite-sample bias, making the clever covariate estimator less stable compared to the weighted regression methods introduced in Section \ref{sec3.4}.

\subsection{Weighted regression estimator}\label{sec3.4}

A third augmented estimator is the weighted regression estimator (WET)\cite{SchaferKang2008, Robins2007, Vansteelandt2011, Gabriel2024} that, in our case, utilizes balancing weights during the outcome model fitting process. Adapted to our setting of survey sampling, the weighted regression estimator can be expressed as,
{\begingroup 
\begin{align} \label{tauWET}
\widehat{\tau}_{h}^{\text{WET}} = \frac{\sum_{i=1}^{n} \displaystyle h(\mathbf{X}_i)/p(\mathbf{X}_i)  \left(m_{1}^{\text{WET}}\left(\mathbf{X}_i; \widehat{\boldsymbol{\alpha}}_{1}^{\text{WET}}\right) -  m_{0}^{\text{WET}}\left(\mathbf{X}_i; \widehat{\boldsymbol{\alpha}}_{0}^{\text{WET}}\right)\right)}{\sum_{i=1}^{n}  h(\mathbf{X}_i)/p(\mathbf{X}_i)}, 
\end{align}
\endgroup}where \(m_{1}^{\text{WET}}\left(\mathbf{X}; \boldsymbol{\alpha}_1^{\text{WET}}\right)\) and \(m_{0}^{\text{WET}}\left(\mathbf{X}; \boldsymbol{\alpha}_0^{\text{WET}}\right)\) are the outcome regression models with canonical link functions, similar to those in the clever covariate estimator,  
but using weighted regression with balancing weights \(\omega_{1,p}\left(\mathbf{X}\right)\) and \(\omega_{0,p}\left(\mathbf{X}\right)\).  
In this estimator, survey weights must be incorporated in three places: estimating the propensity score model $e_{\text{sp}}(\mathbf{X}; \boldsymbol{\beta})$, constructing the balancing weights \(\omega_{z,p}(\mathbf{X}_i)\) that serve as regression weights in the outcome modeling, and the ratio $h(\mathbf{X}_i)/p(\mathbf{X}_i)$ for standardizing predictions to the target population. The weighted regression estimator is asymptotically equivalent to the one-step moment and clever covariate estimators, sharing the same influence function and robustness properties. When $h(\mathbf{X}) \propto 1$, the estimator achieves double robustness. When $h(\mathbf{X})$ depends on $e_{\text{sp}}(\mathbf{X})$, the estimator is singly robust with respect to the propensity score model; if the propensity score model is misspecified, the estimator converges to the shifted estimand $\widetilde{\tau}_{h}$ as discussed in Section \ref{sec3.1}.

Compared to the clever covariate estimator, the weighted regression estimator may offer better performance in scenarios involving highly variable survey weights or extreme propensity scores. As discussed by Kang et al.\cite{Kang2007} and Robins et al.\cite{Robins2007}, the weighted regression estimator is more stable under extreme weights. However, under the IPW scheme, the weighted regression estimator remains susceptible to the impact of extreme inverse propensity scores, which can lead to inflated variance and finite-sample bias, particularly when these extreme weights dominate the regression model. We plan to examine to what extent these observations carry forward to the survey observational study settings in Section \ref{sec4}.

For ease of reference, Table \ref{Table2SummarySWinAllEstimators} summarizes where survey weights need to be applied for each simple weighting estimator and augmented weighting estimator.

\begin{table}[htb]
\caption{Summary of components that incorporate survey weights in propensity score weighting and augmented estimators under the context of retrospective studies using survey observational data.}
\label{Table2SummarySWinAllEstimators}
\setlength{\tabcolsep}{3pt}
\begin{tabular}{p{0.08\textwidth}p{0.17\textwidth}p{0.25\textwidth}p{0.25\textwidth}p{0.17\textwidth}}
\arrayrulecolor{white}
\noalign{\global\arrayrulewidth=0.5pt}
\rowcolor[rgb]{0.92,0.92,0.92}
\textbf{Estimator} & \textbf{Estimating propensity scores} & \textbf{Fitting the outcome regression model} & \textbf{Standardizing outcomes to the target estimand} & \textbf{Standardizing weighted residuals} \\[6pt]

PSW & $\sqrt{}$ & $\times$ & $\sqrt{}$ & $\times$ \\
\rowcolor[rgb]{0.92,0.92,0.92}
MOM & $\sqrt{}$ & $\times$ & $\sqrt{}$ & $\sqrt{}$ \\
CVR & $\sqrt{}$ & $\sqrt{}$ (clever covariates) & $\sqrt{}$ & $\times$ \\
\rowcolor[rgb]{0.92,0.92,0.92}
WET & $\sqrt{}$ & $\sqrt{}$ (weighted regression) & $\sqrt{}$ & $\times$ \\
\end{tabular}

\vspace{1ex}
\raggedright
\footnotesize{Note: All estimators use survey-weighted logistic regression to estimate the population-level propensity score $e_{\text{sp}}(\mathbf{X}; \boldsymbol{\beta})$. PSW uses balancing weights $\omega_{z,p}(\mathbf{X}_i) = h(\mathbf{X}_i)/(p_z(\mathbf{X}_i) e_{\text{sp}}(\mathbf{X}_i))$ to standardize observed outcomes for each group, while augmented estimators use the ratio $h(\mathbf{X}_i)/p(\mathbf{X}_i)$ to standardize predicted full potential outcomes, achieving the target estimand. Additionally, CVR incorporates balancing weights $\omega_{z,p}(\mathbf{X}_i)$ as clever covariates in the outcome model, WET uses $\omega_{z,p}(\mathbf{X}_i)$ as regression weights, and MOM fits an unweighted outcome model but uses $\omega_{z,p}(\mathbf{X}_i)$ to standardize its weighted residuals to achieve additional robustness.}
\end{table}

\subsection{Variance estimation for augmented estimators}\label{sec3.5}

The generalized form of the augmented estimator in Equation \eqref{AWGeneralEstimatorSW} can be expressed as \(\widehat{\tau}_{h}^{\text{AW}} = \widehat{v}_1 + \widehat{v}_2 - \widehat{v}_3\). Leveraging the M-estimation theory, we construct the joint estimating equations \eqref{AWsandwichSW} for sandwich variance estimation.
These estimating equations adjust for the uncertainty in estimating the propensity score and outcome models by utilizing score functions corresponding to \(\boldsymbol{\theta} = \left(v_1, v_2, v_3, \boldsymbol{\alpha}_0^T, \boldsymbol{\alpha}_1^T, \boldsymbol{\beta}_{\text{fp}}^T, \boldsymbol{\beta}_{\text{sp}}^T\right)^T\), 
where \(v_1\), \(v_2\), and \(v_3\) are the components of the general augmented estimator \(\tau_{h}^{\text{AW}}\), \(\boldsymbol{\alpha}_0\) and \(\boldsymbol{\alpha}_1\) are the parameters of the outcome models \(m_0^{\text{AW}}(\mathbf{X}; \boldsymbol{\alpha}_0)\) and \(m_1^{\text{AW}}(\mathbf{X}; \boldsymbol{\alpha}_1)\), respectively, and \(\boldsymbol{\beta}_{\text{fp}}\) and \(\boldsymbol{\beta}_{\text{sp}}\) represent the parameters of the sample-level and population-level propensity score models, respectively.

\begin{equation}\label{AWsandwichSW}
     \mathbf{0}=\sum_{i=1}^n \boldsymbol{\Psi}_{\boldsymbol{\theta}}\left(\mathbf{X}_i, Y_i, Z_i\right)=\sum_{i=1}^n
    \left[
    \begin{array}{c}
        \psi_{\text{PS,FP}} \\
        \psi_{\text{PS,SP}} \\
        \psi_{\text{OR},1} \\
        \psi_{\text{OR},0} \\
        \psi_{\text{v1}} \\
        \psi_{\text{v2}} \\
        \psi_{\text{v3}}
    \end{array}
    \right] 
    = \sum_{i=1}^n\left[
    \begin{array}{l}
    \omega_{\text{PS,FP}}\left(\mathbf{X}_i\right) \mathbf{X}_i \left(Z_i - e_{\text{fp}}\left(\mathbf{X}_i; \boldsymbol{\beta}_{\text{fp}}\right)\right) \\
        \omega_{\text{PS,SP}}\left(\mathbf{X}_i^*\right) \mathbf{X}_i^* \left(Z_i - e_{\text{sp}}\left(\mathbf{X}_i^*; \boldsymbol{\beta}_{\text{sp}}\right)\right) \\
        \omega_{\text{OR},1}\left(\mathbf{X}_i^*\right) Z_i \mathbf{X}_i' \left(Y_i - m_{1}^{\text{AW}}\left(\mathbf{X}_i'; \boldsymbol{\alpha}_1\right)\right) \\
        \omega_{\text{OR},0}\left(\mathbf{X}_i^*\right) (1-Z_i) \mathbf{X}_i'' \left(Y_i - m_{0}^{\text{AW}}\left(\mathbf{X}_i''; \boldsymbol{\alpha}_0\right)\right) \\
        h(\mathbf{X}_i^*)/p(\mathbf{X}_i) \left(m_{1}^{\text{AW}}\left(\mathbf{X}_i'; \boldsymbol{\alpha}_1\right) - m_{0}^{\text{AW}}\left(\mathbf{X}_i''; \boldsymbol{\alpha}_0\right) - v_1\right) \\
        \omega_{1,p}(\mathbf{X}_i^*) Z_i  \left(Y_i - m_{1}^{\text{AW}}\left(\mathbf{X}_i'; \boldsymbol{\alpha}_1\right) - v_2\right) \\
       \omega_{0,p}(\mathbf{X}_i^*)  (1-Z_i) \left(Y_i - m_{0}^{\text{AW}}\left(\mathbf{X}_i''; \boldsymbol{\alpha}_0\right) - v_3\right)
    \end{array}
    \right].
\end{equation}

The equation above includes seven distinct components: \(\psi_{\text{PS,FP}}\), \(\psi_{\text{PS,SP}}\), \(\psi_{\text{OR},1}\), \(\psi_{\text{OR},0}\), \(\psi_{\text{v1}}\), \(\psi_{\text{v2}}\), and \(\psi_{\text{v3}}\), each representing different aspects of the augmented weighting estimation process. The first two components, \(\psi_{\text{PS,FP}}\) and \(\psi_{\text{PS,SP}}\), correspond to the score functions associated with the estimation of the propensity score model at the sample and population levels, respectively. 
In retrospective settings, \(e_{\text{fp}}(\mathbf{X}; \boldsymbol{\beta}_{\text{fp}})\) is used exclusively to calculate \(r_z(\mathbf{X}_i)\) for obtaining the marginalized sampling probability \(p(\mathbf{X}_i) = r_z(\mathbf{X}_i) p_z(\mathbf{X}_i)\), which is used for tilting the imputed outcome difference in the augmented estimator as discussed in Section \ref{sec3.1}.
The term \(\omega_{\text{PS,FP}}(\mathbf{X}_i)\) represents the regression weights for estimating the within-sample propensity score and is set to \(1\). 
Furthermore, \(\omega_{\text{PS,SP}}(\mathbf{X}_i^*)\) refers to the regression weights applied to the covariate vector \(\mathbf{X}_i^*\) in the population-level propensity score model \(e_{\text{sp}}(\mathbf{X}; \boldsymbol{\beta}_{\text{sp}})\), which is either the survey weights when the regression is survey-weighted (W.PS) or set to \(1\) when the regression is unweighted (U.PS).
The covariate vector \(\mathbf{X}_i^*\) may consist of baseline covariates alone or may additionaly include survey weights as a covariate (C.PS), as recommended by DuGoff et al. \cite{DuGoff2014} to capture unique design features and address latent sampling information so as to enhance the covariate balance.

The components \(\psi_{\text{OR},1}\) and \(\psi_{\text{OR},0}\) represent the score functions for the outcome models for the treated (\(z_i = 1\)) and control (\(z_i = 0\)) groups, respectively. The regression weights \(\omega_{\text{OR},1}(\mathbf{X}_i^*)\) and \(\omega_{\text{OR},0}(\mathbf{X}_i^*)\) depend on the choice of augmented estimator: for the one-step moment estimator (MOM) and clever covariate estimator (CVR), these weights are set to 1; for the weighted regression estimator (WET), they correspond to the balancing weights \(\omega_{1,p}(\mathbf{X}_i^*)\) and \(\omega_{0,p}(\mathbf{X}_i^*)\) as specified in Section \ref{sec3.4}. The outcome models \(m_{1}^{\text{AW}}\) and \(m_{0}^{\text{AW}}\) reflect the specific type of augmentation used, where the superscript “AW” can denote "MOM", "CVR", or "WET". 
For the clever covariate estimator, the outcome model covariates \(\mathbf{X}_i'\) and \(\mathbf{X}_i''\) include the clever covariates for each arm, whereas for MOM and WET, only the baseline covariates are used, i.e., \(\mathbf{X}_i\).

The component \(\psi_{v_1}\) corresponds to the first component \(v_1\) in the general augmented treatment effect estimator \eqref{AWGeneralEstimatorSW}, representing the difference in predicted potential outcomes between the treated and control groups, adjusted by the ratio \(h(\mathbf{X}_i^*)/p(\mathbf{X}_i)\), where the tilting function $h(\mathbf{X}_i^*)$ may depend on $e_{\text{sp}}(\mathbf{X}_i^*; \boldsymbol{\beta}_{\text{sp}})$ as specified in Table \ref{Table1CommonBalancingWeightsUnderSurveySettings}.
The components \(\psi_{v_2}\) and \(\psi_{v_3}\) represent the residual adjustments for the treated and control groups, respectively, incorporating the balancing weights \(\omega_{1,p}(\mathbf{X}_i^*)\) and \(\omega_{0,p}(\mathbf{X}_i^*)\) to adjust for confounding, as described in Table \ref{Table1CommonBalancingWeightsUnderSurveySettings}.

These seven components together form the complete joint estimating equations, 
capturing the uncertainties associated with the estimated propensity score, outcome models, and treatment effects in a comprehensive manner. Under suitable regularity conditions,\cite{stefanski2002calculus} each type of augmented estimator achieves asymptotic normality \(\sqrt{n}(\widehat{\boldsymbol{\theta}} - \boldsymbol{\theta}) \xrightarrow{d} N(\mathbf{0}, \mathbf{V}(\boldsymbol{\theta}))\), where the variance-covariance matrix \(\mathbf{V}(\boldsymbol{\theta})\) is defined as \(\mathbf{V}(\boldsymbol{\theta}) = \mathbf{A}(\boldsymbol{\theta})^{-1} \mathbf{B}(\boldsymbol{\theta}) \{\mathbf{A}(\boldsymbol{\theta})^{-1}\}^T\) with \(\mathbf{A}(\boldsymbol{\theta}) = E\left\{-\displaystyle\frac{\partial  \boldsymbol{\Psi}_{\boldsymbol{\theta}}\left(\mathbf{X}, Y, Z\right)}{\partial \boldsymbol{\theta}} \mid S=1\right\}\), and \(\mathbf{B}(\boldsymbol{\theta}) = E\left\{ \boldsymbol{\Psi}_{\boldsymbol{\theta}}\left(\mathbf{X}, Y, Z\right)  \boldsymbol{\Psi}_{\boldsymbol{\theta}}\left(\mathbf{X}, Y, Z\right)^T \mid S=1 \right\}\). The quantities \(\mathbf{A}(\boldsymbol{\theta})\) and \(\mathbf{B}(\boldsymbol{\theta}) \) can be consistently estimated by replacing the expected values with their empirical counterparts, \(\widehat{\mathbf{A}}(\widehat{\boldsymbol{\theta}}) = -\frac{1}{n}\sum_{i=1}^n {\partial  \boldsymbol{\Psi}_{\widehat{\boldsymbol{\theta}}}\left(\mathbf{X}_i, Y_i, Z_i\right)}/{\partial \boldsymbol{\theta}}\) and \(\widehat{\mathbf{B}}(\widehat{\boldsymbol{\theta}}) = \frac{1}{n}\sum_{i=1}^n  \boldsymbol{\Psi}_{\widehat{\boldsymbol{\theta}}}\left(\mathbf{X}_i, Y_i, Z_i\right)  \boldsymbol{\Psi}_{\widehat{\boldsymbol{\theta}}}\left(\mathbf{X}_i, Y_i, Z_i\right)^T\)
, resulting in the empirical sandwich variance estimator \(\widehat{\mathbf{V}}(\widehat{\boldsymbol{\theta}}) = \widehat{\mathbf{A}}(\widehat{\boldsymbol{\theta}})^{-1} \widehat{\mathbf{B}}(\widehat{\boldsymbol{\theta}}) \{\widehat{\mathbf{A}}(\widehat{\boldsymbol{\theta}})^{-1}\}^T\). The variance of the augmented estimator \(\widehat{\tau}_{h}^{\text{AW}}\) can be consistently estimated using the corresponding elements of \(\widehat{\mathbf{V}}(\widehat{\boldsymbol{\theta}})\).

\begin{table}[htb]
\captionsetup{justification=raggedright} 
\caption{Covariate vector notations used in the estimating equations for the sandwich variance estimation of augmented estimators for survey observational data.}
\label{table:notations}
\setlength{\tabcolsep}{8pt}
\begin{tabular}{p{0.08\textwidth}p{0.84\textwidth}}
\arrayrulecolor{white}
\noalign{\global\arrayrulewidth=0.5pt}
\rowcolor[rgb]{0.92,0.92,0.92}
\textbf{Notation} & \textbf{Definition and Context} \\[6pt]

\( \mathbf{X}_i^* \) & Covariates used in the propensity score model \( e_{\text{sp}}(\mathbf{X}_i^*; \boldsymbol{\beta}_{\text{sp}}) \). For unweighted (U.PS) and survey-weighted regression scenarios (W.PS), it includes only baseline covariates. When survey weights are incorporated as an additional covariate (C.PS), it contains both the baseline covariates and the survey weights. \\

\rowcolor[rgb]{0.92,0.92,0.92}
\( \mathbf{X}_i' \) & Covariates in the treated outcome regression model \( m_{1}^{\text{AW}}(\mathbf{X}_i'; \boldsymbol{\alpha}_1) \). For CVR, it includes clever covariates; for MOM and WET, only baseline covariates are used. \\

\( \mathbf{X}_i'' \) & Covariates in the control outcome regression model \( m_{0}^{\text{AW}}(\mathbf{X}_i''; \boldsymbol{\alpha}_0) \). Defined analogously to \( \mathbf{X}_i' \) based on the estimator type. \\
\end{tabular}

\vspace{1ex}
\raggedright
\footnotesize{Note: The specified original covariates used in the propensity score model may differ from those in the outcome regression models, and the covariates used in the outcome models for each arm are not necessarily the same. In MOM and WET, where both outcome models include only baseline covariates without distinct clever covariates, \(\mathbf{X}_i'\) and \(\mathbf{X}_i''\) may be identical or even unified, using one model for both arms' counterfactual outcome predictions. In cases where the same covariates are used across all models, such as in MOM and WET with U.PS or W.PS, all three covariate notations (\(\mathbf{X}_i^*\), \(\mathbf{X}_i'\), and \(\mathbf{X}_i''\)) could be identical.}
\end{table}

\section{Simulation Studies}\label{sec4}

\subsection{Simulation design}\label{sec4.1}

Our data-generating process was adapted from the multi-stage strategy employed by Austin et al. and Lenis et al. \cite{AustinJembereChiu2018,Lenis2019} We simulated a population of $1,000,000$ individuals, organized into 10 strata \((j=1, \ldots, 10)\). Each stratum was further divided into 20 clusters, resulting in a total of 200 clusters indexed globally as \( (k=1, \ldots, 200) \), with  5000 individuals in each cluster \((i=1, \ldots, 5000)\). For each individual, we generated six baseline covariates \( (l=1, \ldots, 6) \) from normal distributions, where the vector of covariates for the \(i\)-th individual is denoted as \(\mathbf{X}_i = \{X_{1i}, X_{2i}, X_{3i}, X_{4i}, X_{5i}, X_{6i}\}\). The means of these covariates were allowed to vary at both the stratum and cluster levels. For the $l$-th covariate, a stratum-specific random effect for the $j$ th stratum $\nu_{l, j}^{\text {stratum }} \sim N\left(0, \sigma_l^{\text {stratum }}\right)$ was generated, along with a cluster-specific random effect for the $k$ -th cluster $\nu_{l, k}^{\text {cluster }} \sim N\left(0, \sigma_l^{\text {cluster }}\right)$. 
The $l$-th covariate for the $i$-th individual in the $j$-th stratum and $k$ -th cluster was then generated as $X_{l, i j k} \sim N\left(\nu_{l, j}^{\text {stratum }}+\nu_{l, k}^{\text {cluster }}, 1\right)$. This hierarchical structure builds in correlation within clusters (through shared stratum and cluster effects) as well as within strata (through shared stratum effects only), while maintaining independence between strata.
We set \(\sigma^{\text {stratum }} = 0.35\) and \(\sigma^{\text {cluster }} = 0.15\) throughout.

We specified the true population-level propensity score for each individual using a logistic model:  
\begin{align}
    \operatorname{logit}\left(e_{\text{sp}}(\mathbf{X}_i)\right) =  a_0 + \psi \left(\boldsymbol{a}_{1:6}^T \mathbf{X}_{i} + a_7 X_{1i} X_{2i}\right),\nonumber
\end{align}
where \(\boldsymbol{a}_{1:6} = (\log(1.1), \log(1.25), \log(1.5), \log(1.75), \log(2), \log(2.5))^T\), and \(a_7 = \log(1.1)\). Treatment assignment \(Z_i\) was simulated from a Bernoulli distribution: \( Z_i \sim \operatorname{Bernoulli}(e_{\text{sp}}(\mathbf{X}_i)) \). The scalar \(\psi\) reflects the strength of confounding and dictates the level of overlap in the propensity score. We set \(\psi = 0.6\) or \(2\) to mimic scenarios with good or poor overlap in the full population, as visualized in Web Figure 1(a) and Web Figure 1(b), respectively. For both scenarios, the overall treatment prevalence was maintained at approximately 30\% by adjusting \(a_0\) to \(\log (35/80)\) for the good overlap and \(\log(20/80)\) for the poor overlap, allowing us to isolate the impact of overlap when evaluating estimator performance.

We simulated continuous outcomes for each individual \(i\) in the population using a linear combination of covariates, treatment effects, and interaction terms, with
\begin{align}
    Y_i = b_0 + \delta_0 \left(\mathbf{b}^T \mathbf{X}_i + b_7 X_{1i} X_{2i}\right) + Z_i \left( \delta_1 + \delta_2 \left(\mathbf{b}^T \mathbf{X}_i + b_8 X_{1i} X_{2i}\right) \right) + \varepsilon, \nonumber
\end{align}
where \(\varepsilon \sim N(0,1)\). We set \(b_0 = 0\) and \(\boldsymbol{b} = (2.5, -2, 1.75, -1.25, 1.5, 1.1)^T\). The average treatment effect was represented by \(\delta_1 = 1\), while \(\delta_0 = 0.3\) and \(\delta_2 = 0.2\) introduced heterogeneity in the control and treatment conditions, respectively. Additionally, two interaction terms were included, with \(b_7 = 2.5\) and \(b_8 = 1.5\), to capture the more complex influence of baseline covariates on outcomes across both conditions.

We drew 1,000 random samples with 5000 individuals each  (representing 0.5\% of the overall population) first under a prospective framework using a multi-stage sampling design consistent with Austin et al.\cite{AustinJembereChiu2018} and Lenis et al.,\cite{Lenis2019}, where the sampling mechanism was independent of the treatment assignment mechanism. Sample sizes were allocated across the 10 strata as \(\{850, 750, 700, 650, 600, 400, 350, 300, 250, 150\}\), ensuring that the observed sample would have a different distribution from the population. Within each stratum, five clusters were selected using simple random sampling, and from each selected cluster, an equal number of individuals were sampled. For each sampled individual, we calculated the true survey weights by taking the inverse of the sampling probability, reflecting the number of individuals in the population that the sampled individual represents.

To further assess the performance of our estimators under retrospective designs, we extended the simulation by introducing an alternative sampling mechanism which now depends on both baseline covariates and treatment assignment. We used logistic regression to model the sampling probability for individual \(i\) as a function of both the treatment assignment indicator \(Z_i\) and covariates \(\mathbf{X}_i\), allowing sampling to reflect more complex real-world dependencies. The sampling probability \(p_{z,i} = P(S_i = 1 \mid Z_i , \mathbf{X}_i ) \) was computed as \(\operatorname{logit}(p_{z,i}) = c_0 + \delta_z^{(s)}  Z_i + \mathbf{c}^T \mathbf{X}_i\), where \( c_0 = \log\left(\frac{0.005}{1 - 0.005}\right) \) defines the baseline sampling rate of 0.5\% consistent with the above multi-stage design, and \( \delta_z^{(s)} = \log(0.9) \) slightly reduces the sampling probability for the treatment group. The coefficient vector was specified as \( \mathbf{c} = (c_1, c_2, c_3, c_4, c_5, c_6)^T=(\log(1.05), \log(1.10), \log(1.15), \log(1.10), \log(1.05), \log(1.10))^T \).
The sampling indicator \( S_i \) was subsequently generated as a Bernoulli random variable with probability \( p_{z,i}\) for individual $i$, capturing the interaction between treatment assignment and covariates in determining sample inclusion. More details and results of this alternative simulation are presented in Web Appendix B.2.

For each sample within each scenario, we evaluated four types of estimators: the PSW (propensity score weighting estimator) and three augmentation methods, referred to as MOM, CVR and WET. To investigate the role of survey weights in estimating the propensity scores, we primarily compare two strategies for integrating survey weights in the propensity score model: as model weights (W.PS) and as an additional covariate (C.PS). The latter involves directly adjusting for survey weight as one additional single covariate
in the propensity score model, which is distinct from including the balancing weight as a clever covariate in the outcome regression model. Additionally, results for unweighted propensity scores (U.PS) and survey weights applied as both model weights and covariates (CW.PS) are presented in the Web Appendix B.1.
These comparisons may shed light on previous debates in the literature regarding which level of covariate balance should we achieve with propensity score and in what way.\cite{DuGoff2014, Ridgeway2015}
For estimating the sample-level propensity score $e_{\text{fp}}(\mathbf{X})$ used in approximating $p(\mathbf{X}_i)$ for the augmented estimators, we use standard unweighted logistic regression as an effective approximation under this rare sampling condition for the theoretically correct nonlinear logistic form in Equation \eqref{eq:efp-retro-ps}). See Web Appendix B.3.1 for additional sensitivity analyses.

In total, we constructed eight estimators by combining the four types of estimators with the two methods of incorporating survey weights into the propensity score model. These estimators were evaluated across three common estimands: the population average treatment effect (PATE), the population average treatment effect for the treated, (PATT), and the population average treatment effect among the overlap population (PATO). Additionally, we assessed their performance under four model misspecification scenarios: (i) correct propensity score model and correct outcome model (Cor\textbar Cor), (ii) incorrect propensity score model but correct outcome model (Mis\textbar Cor), (iii) correct propensity score model but incorrect outcome model (Cor\textbar Mis), and (iv) both models misspecified (Mis\textbar Mis).

Given the complexity of the design, it is instructive to note the total number of considered scenarios. We examined two levels of overlap (good vs. poor), two primary methods for incorporating survey weights into the propensity score model (W.PS vs. C.PS), four estimators (PSW, MOM, CVR, WET), three estimands (PATE, PATT, PATO), and four model specification scenarios (Cor\textbar Cor, Mis\textbar Cor, Cor\textbar Mis, Mis\textbar Mis). This combination yields $2 \times 2 \times 4 \times 3 \times 4 = 192$ distinct configurations in the main analysis. Additional scenarios that incorporate the unweighted and combined weighting-plus-covariate propensity score approaches and the sampling mechanism depending on treatment increase this number even further. The factorial nature of this design ensures that our conclusions are drawn from a broad range of realistic conditions, and the resulting comparisons are adequately comprehensive. Table \ref{tab:scenario_counts} outlines these factors and their levels for a quick reference.



\begin{table}[!htbp]
\caption{Summary of factors and number of scenarios in the main simulation design}
\label{tab:scenario_counts}
\centering
\setlength{\tabcolsep}{6pt} 
\renewcommand{\arraystretch}{1.1} 
\begin{tabularx}{1.05\textwidth}{l X c} 
\arrayrulecolor{white}
\rowcolor[rgb]{0.92,0.92,0.92}
\textbf{Factor} & \textbf{Levels} & \textbf{Count} \\[6pt]
\arrayrulecolor{black}\hline
Overlap scenario & Good vs. Poor & 2 \\
\rowcolor[rgb]{0.92,0.92,0.92}
Method of incorporating survey weights into PS model & W.PS, C.PS & 2 \\
Estimator & PSW, MOM, CVR, WET & 4 \\
\rowcolor[rgb]{0.92,0.92,0.92}
Estimand & PATE, PATT, PATO & 3 \\
Model specification scenario & Cor\textbar Cor, Mis\textbar Cor, Cor\textbar Mis, Mis\textbar Mis & 4 \\
\hline
\rowcolor[rgb]{0.92,0.92,0.92}
Total scenarios in main analysis & & 192 \\
\bottomrule
\end{tabularx}
\end{table}

Beyond the main simulation framework, we conducted sensitivity analyses to assess the robustness of our proposed estimators under alternative configurations. In Web Appendix B.3, we systematically evaluate two aspects: (1) the choice between the principled approach using standard unweighted logistic regression versus maximum likelihood estimation for the theoretically correct nonlinear logistic form of the sample-level propensity score $e_{\text{fp}}(\mathbf{X})$ under retrospective treatment-dependent sampling (Section B.3.1), and (2) robustness across diverse sampling mechanisms including varying treatment-dependent sampling strengths (odds ratios of $p_1(\mathbf{x})/p_0(\mathbf{x})$ from 0.5 to 2.0), complex $Z$-$X$ interactions in the true sampling model, and alternative link functions such as probit (Section B.3.2). These sensitivity analyses demonstrate the robust performance of our proposed estimators under possible model misspecification and varying complex sampling designs.

Performance metrics included relative bias, relative efficiency, and empirical coverage of the 95\% confidence intervals. Relative bias was calculated as the percentage difference between the estimated and true population treatment effects, averaged across 1,000 simulations. Relative efficiency was determined as the ratio of the Monte Carlo variance of the reference estimator (the balancing weight estimator for PATE with W.PS under correct model specification) to that of each evaluated estimator. Empirical coverage was assessed by calculating the proportion of confidence intervals, constructed using sandwich variance estimators, that contained the true population treatment effect. The simulation code and implementations of the survey-weighted balancing weight estimator and the augmented estimators for PATE, PATT, and PATO are available online at \url{https://github.com/ykzeng-yale/SW_PSW}, and they are implemented in the PSweight R package (Version 2.1.0).

\subsection{Simulations results}\label{sec4.2}

In this section, we focus on the primary four types of estimators (PSW, MOM, CVR and WET) and two distinct methods of incorporating survey weights into propensity score models (W.PS and C.PS). Table \ref{GoodOverlapSimulationResults} presents the simulation results under the good overlap scenario, summarizing the relative bias, relative efficiency, and coverage of PATE, PATT, and PATO under various modeling strategies with either correctly or incorrectly specified propensity score and outcome regression models. In parallel, Table \ref{PoorOverlapSimulationResults} presents the results when the overlap is poor. 
In addition to these primary results, we present more comprehensive analyses that include U.PS and CW.PS in Web Tables 3 to 5 and 6 to 8 for each estimand under good and poor overlap scenarios. 
Furthermore, simulation results for the extended scenario, where sampling depends on treatment assignment, are provided in Web Tables 9 and 10 for W.PS and C.PS results, and Web Tables 11 to 13 and 14 to 16 for more comprehensive analyses.
To facilitate the summary of findings, we focus on the following three thorny questions; that is, how survey weights should be incorporated at each stage of the balancing weight estimator, which augmented estimators perform better, and which method is empirically more robust (i.e., under poor overlap and model misspecification).

\begin{sidewaystable}[htbp]
\centering
\caption{Relative bias (\%), relative efficiency, and coverage of the 95\% confidence intervals for the estimators for continuous outcomes under good overlap across 5000 simulations.} \label{GoodOverlapSimulationResults}
\label{ExtractedGoodOverlapSimulationResults}
\scriptsize
\renewcommand{\arraystretch}{0.8}
\setlength{\tabcolsep}{0.5pt}
\begin{tabularx}{\textwidth}{@{}>{\centering\arraybackslash}p{1.5cm} *{6}{>{\centering\arraybackslash}p{1.2cm}} *{12}{>{\centering\arraybackslash}p{1.2cm}} @{}}
\toprule
\multirow{2}{*}{Estimand} & \multicolumn{6}{c}{Specification} & \multicolumn{4}{c}{Relative Bias(\%)} & \multicolumn{4}{c}{Relative Efficiency} & \multicolumn{4}{c}{Coverage} \\
\cmidrule(lr){2-7} \cmidrule(lr){8-11} \cmidrule(lr){12-15} \cmidrule(lr){16-19}
& W.PS & C.PS & PSW & MOM & CVR & WEI & Cor\textbar Cor & Mis\textbar Cor & Cor\textbar Mis & Mis\textbar Mis & Cor\textbar Cor & Mis\textbar Cor & Cor\textbar Mis & Mis\textbar Mis & Cor\textbar Cor & Mis\textbar Cor & Cor\textbar Mis & Mis\textbar Mis \\
\midrule
\multirow{8}{*}{PATE}
& \checkmark &  & \checkmark &  &  &  & 0.554 & 8.742 & \textendash & \textendash & 1.000 & 0.757 & \textendash & \textendash & 0.951 & 0.795 & \textendash & \textendash \\[2pt]
& \cellcolor[rgb]{0.92,0.92,0.92} & \cellcolor[rgb]{0.92,0.92,0.92}\checkmark & \cellcolor[rgb]{0.92,0.92,0.92}\checkmark & \cellcolor[rgb]{0.92,0.92,0.92} & \cellcolor[rgb]{0.92,0.92,0.92} & \cellcolor[rgb]{0.92,0.92,0.92} & \cellcolor[rgb]{0.92,0.92,0.92}0.741 & \cellcolor[rgb]{0.92,0.92,0.92}9.144 & \cellcolor[rgb]{0.92,0.92,0.92}\textendash & \cellcolor[rgb]{0.92,0.92,0.92}\textendash & \cellcolor[rgb]{0.92,0.92,0.92}0.729 & \cellcolor[rgb]{0.92,0.92,0.92}0.621 & \cellcolor[rgb]{0.92,0.92,0.92}\textendash & \cellcolor[rgb]{0.92,0.92,0.92}\textendash & \cellcolor[rgb]{0.92,0.92,0.92}0.926 & \cellcolor[rgb]{0.92,0.92,0.92}0.786 & \cellcolor[rgb]{0.92,0.92,0.92}\textendash & \cellcolor[rgb]{0.92,0.92,0.92}\textendash \\[2pt]
& \checkmark &  &  & \checkmark &  &  & 0.188 & 0.211 & 0.236 & 8.704 & 1.521 & 1.521 & 1.365 & 0.900 & 0.929 & 0.929 & 0.983 & 0.796 \\[2pt]
& \cellcolor[rgb]{0.92,0.92,0.92} & \cellcolor[rgb]{0.92,0.92,0.92}\checkmark & \cellcolor[rgb]{0.92,0.92,0.92} & \cellcolor[rgb]{0.92,0.92,0.92}\checkmark & \cellcolor[rgb]{0.92,0.92,0.92} & \cellcolor[rgb]{0.92,0.92,0.92} & \cellcolor[rgb]{0.92,0.92,0.92}0.171 & \cellcolor[rgb]{0.92,0.92,0.92}0.194 & \cellcolor[rgb]{0.92,0.92,0.92}0.218 & \cellcolor[rgb]{0.92,0.92,0.92}8.766 & \cellcolor[rgb]{0.92,0.92,0.92}1.520 & \cellcolor[rgb]{0.92,0.92,0.92}1.519 & \cellcolor[rgb]{0.92,0.92,0.92}1.201 & \cellcolor[rgb]{0.92,0.92,0.92}0.893 & \cellcolor[rgb]{0.92,0.92,0.92}0.930 & \cellcolor[rgb]{0.92,0.92,0.92}0.930 & \cellcolor[rgb]{0.92,0.92,0.92}0.974 & \cellcolor[rgb]{0.92,0.92,0.92}0.795 \\[2pt]
& \checkmark &  &  &  & \checkmark &  & 0.214 & 0.196 & 8.152 & 6.791 & 1.395 & 1.394 & 0.750 & 0.809 & 0.940 & 0.940 & 0.854 & 0.859 \\[2pt]
& \cellcolor[rgb]{0.92,0.92,0.92} & \cellcolor[rgb]{0.92,0.92,0.92}\checkmark & \cellcolor[rgb]{0.92,0.92,0.92} & \cellcolor[rgb]{0.92,0.92,0.92} & \cellcolor[rgb]{0.92,0.92,0.92}\checkmark & \cellcolor[rgb]{0.92,0.92,0.92} & \cellcolor[rgb]{0.92,0.92,0.92}0.195 & \cellcolor[rgb]{0.92,0.92,0.92}0.174 & \cellcolor[rgb]{0.92,0.92,0.92}8.152 & \cellcolor[rgb]{0.92,0.92,0.92}6.677 & \cellcolor[rgb]{0.92,0.92,0.92}1.393 & \cellcolor[rgb]{0.92,0.92,0.92}1.392 & \cellcolor[rgb]{0.92,0.92,0.92}0.739 & \cellcolor[rgb]{0.92,0.92,0.92}0.810 & \cellcolor[rgb]{0.92,0.92,0.92}0.940 & \cellcolor[rgb]{0.92,0.92,0.92}0.941 & \cellcolor[rgb]{0.92,0.92,0.92}0.848 & \cellcolor[rgb]{0.92,0.92,0.92}0.860 \\[2pt]
& \checkmark &  &  &  &  & \checkmark & 0.188 & 0.196 & 0.272 & 8.596 & 1.525 & 1.525 & 1.380 & 0.919 & 0.935 & 0.935 & 0.986 & 0.804 \\[2pt]
& \cellcolor[rgb]{0.92,0.92,0.92} & \cellcolor[rgb]{0.92,0.92,0.92}\checkmark & \cellcolor[rgb]{0.92,0.92,0.92} & \cellcolor[rgb]{0.92,0.92,0.92} & \cellcolor[rgb]{0.92,0.92,0.92} & \cellcolor[rgb]{0.92,0.92,0.92}\checkmark & \cellcolor[rgb]{0.92,0.92,0.92}0.169 & \cellcolor[rgb]{0.92,0.92,0.92}0.177 & \cellcolor[rgb]{0.92,0.92,0.92}0.251 & \cellcolor[rgb]{0.92,0.92,0.92}8.581 & \cellcolor[rgb]{0.92,0.92,0.92}1.525 & \cellcolor[rgb]{0.92,0.92,0.92}1.524 & \cellcolor[rgb]{0.92,0.92,0.92}1.219 & \cellcolor[rgb]{0.92,0.92,0.92}0.918 & \cellcolor[rgb]{0.92,0.92,0.92}0.935 & \cellcolor[rgb]{0.92,0.92,0.92}0.935 & \cellcolor[rgb]{0.92,0.92,0.92}0.975 & \cellcolor[rgb]{0.92,0.92,0.92}0.805 \\
\midrule
\multirow{8}{*}{PATT}
& \checkmark &  & \checkmark &  &  &  & -0.222 & 5.515 & \textendash & \textendash & 1.220 & 0.882 & \textendash & \textendash & 0.937 & 0.805 & \textendash & \textendash \\[2pt]
& \cellcolor[rgb]{0.92,0.92,0.92} & \cellcolor[rgb]{0.92,0.92,0.92}\checkmark & \cellcolor[rgb]{0.92,0.92,0.92}\checkmark & \cellcolor[rgb]{0.92,0.92,0.92} & \cellcolor[rgb]{0.92,0.92,0.92} & \cellcolor[rgb]{0.92,0.92,0.92} & \cellcolor[rgb]{0.92,0.92,0.92}-0.034 & \cellcolor[rgb]{0.92,0.92,0.92}5.792 & \cellcolor[rgb]{0.92,0.92,0.92}\textendash & \cellcolor[rgb]{0.92,0.92,0.92}\textendash & \cellcolor[rgb]{0.92,0.92,0.92}0.886 & \cellcolor[rgb]{0.92,0.92,0.92}0.725 & \cellcolor[rgb]{0.92,0.92,0.92}\textendash & \cellcolor[rgb]{0.92,0.92,0.92}\textendash & \cellcolor[rgb]{0.92,0.92,0.92}0.898 & \cellcolor[rgb]{0.92,0.92,0.92}0.773 & \cellcolor[rgb]{0.92,0.92,0.92}\textendash & \cellcolor[rgb]{0.92,0.92,0.92}\textendash \\[2pt]
& \checkmark &  &  & \checkmark &  &  & -0.035 & -1.358 & -0.175 & 5.436 & 1.425 & 1.456 & 1.368 & 0.954 & 0.913 & 0.907 & 0.935 & 0.829 \\[2pt]
& \cellcolor[rgb]{0.92,0.92,0.92} & \cellcolor[rgb]{0.92,0.92,0.92}\checkmark & \cellcolor[rgb]{0.92,0.92,0.92} & \cellcolor[rgb]{0.92,0.92,0.92}\checkmark & \cellcolor[rgb]{0.92,0.92,0.92} & \cellcolor[rgb]{0.92,0.92,0.92} & \cellcolor[rgb]{0.92,0.92,0.92}-0.098 & \cellcolor[rgb]{0.92,0.92,0.92}-1.471 & \cellcolor[rgb]{0.92,0.92,0.92}-0.226 & \cellcolor[rgb]{0.92,0.92,0.92}5.346 & \cellcolor[rgb]{0.92,0.92,0.92}1.469 & \cellcolor[rgb]{0.92,0.92,0.92}1.491 & \cellcolor[rgb]{0.92,0.92,0.92}1.268 & \cellcolor[rgb]{0.92,0.92,0.92}0.974 & \cellcolor[rgb]{0.92,0.92,0.92}0.911 & \cellcolor[rgb]{0.92,0.92,0.92}0.902 & \cellcolor[rgb]{0.92,0.92,0.92}0.923 & \cellcolor[rgb]{0.92,0.92,0.92}0.828 \\[2pt]
& \checkmark &  &  &  & \checkmark &  & -0.067 & -1.390 & 1.778 & 6.163 & 1.649 & 1.687 & 1.470 & 1.114 & 0.936 & 0.931 & 0.931 & 0.826 \\[2pt]
& \cellcolor[rgb]{0.92,0.92,0.92} & \cellcolor[rgb]{0.92,0.92,0.92}\checkmark & \cellcolor[rgb]{0.92,0.92,0.92} & \cellcolor[rgb]{0.92,0.92,0.92} & \cellcolor[rgb]{0.92,0.92,0.92}\checkmark & \cellcolor[rgb]{0.92,0.92,0.92} & \cellcolor[rgb]{0.92,0.92,0.92}-0.128 & \cellcolor[rgb]{0.92,0.92,0.92}-1.502 & \cellcolor[rgb]{0.92,0.92,0.92}1.696 & \cellcolor[rgb]{0.92,0.92,0.92}6.062 & \cellcolor[rgb]{0.92,0.92,0.92}1.711 & \cellcolor[rgb]{0.92,0.92,0.92}1.736 & \cellcolor[rgb]{0.92,0.92,0.92}1.478 & \cellcolor[rgb]{0.92,0.92,0.92}1.139 & \cellcolor[rgb]{0.92,0.92,0.92}0.938 & \cellcolor[rgb]{0.92,0.92,0.92}0.939 & \cellcolor[rgb]{0.92,0.92,0.92}0.826 & \cellcolor[rgb]{0.92,0.92,0.92}0.826 \\[2pt]
& \checkmark &  &  &  &  & \checkmark & -0.034 & -1.360 & -0.128 & 5.423 & 1.427 & 1.457 & 1.374 & 0.959 & 0.940 & 0.932 & 0.863 & 0.863 \\[2pt]
& \cellcolor[rgb]{0.92,0.92,0.92} & \cellcolor[rgb]{0.92,0.92,0.92}\checkmark & \cellcolor[rgb]{0.92,0.92,0.92} & \cellcolor[rgb]{0.92,0.92,0.92} & \cellcolor[rgb]{0.92,0.92,0.92} & \cellcolor[rgb]{0.92,0.92,0.92}\checkmark & \cellcolor[rgb]{0.92,0.92,0.92}-0.099 & \cellcolor[rgb]{0.92,0.92,0.92}-1.475 & \cellcolor[rgb]{0.92,0.92,0.92}-0.181 & \cellcolor[rgb]{0.92,0.92,0.92}5.319 & \cellcolor[rgb]{0.92,0.92,0.92}1.473 & \cellcolor[rgb]{0.92,0.92,0.92}1.494 & \cellcolor[rgb]{0.92,0.92,0.92}1.274 & \cellcolor[rgb]{0.92,0.92,0.92}0.980 & \cellcolor[rgb]{0.92,0.92,0.92}0.938 & \cellcolor[rgb]{0.92,0.92,0.92}0.927 & \cellcolor[rgb]{0.92,0.92,0.92}0.864 & \cellcolor[rgb]{0.92,0.92,0.92}0.864 \\
\midrule
\multirow{8}{*}{PATO}
& \checkmark &  & \checkmark &  &  &  & 0.328 & 7.286 & \textendash & \textendash & 1.633 & 1.049 & \textendash & \textendash & 0.951 & 0.766 & \textendash & \textendash \\[2pt]
& \cellcolor[rgb]{0.92,0.92,0.92} & \cellcolor[rgb]{0.92,0.92,0.92}\checkmark & \cellcolor[rgb]{0.92,0.92,0.92}\checkmark & \cellcolor[rgb]{0.92,0.92,0.92} & \cellcolor[rgb]{0.92,0.92,0.92} & \cellcolor[rgb]{0.92,0.92,0.92} & \cellcolor[rgb]{0.92,0.92,0.92}0.481 & \cellcolor[rgb]{0.92,0.92,0.92}7.602 & \cellcolor[rgb]{0.92,0.92,0.92}\textendash & \cellcolor[rgb]{0.92,0.92,0.92}\textendash & \cellcolor[rgb]{0.92,0.92,0.92}1.042 & \cellcolor[rgb]{0.92,0.92,0.92}0.817 & \cellcolor[rgb]{0.92,0.92,0.92}\textendash & \cellcolor[rgb]{0.92,0.92,0.92}\textendash & \cellcolor[rgb]{0.92,0.92,0.92}0.914 & \cellcolor[rgb]{0.92,0.92,0.92}0.755 & \cellcolor[rgb]{0.92,0.92,0.92}\textendash & \cellcolor[rgb]{0.92,0.92,0.92}\textendash \\[2pt]
& \checkmark &  &  & \checkmark &  &  & 0.286 & -0.306 & 0.299 & 7.322 & 1.648 & 1.664 & 1.646 & 1.049 & 0.925 & 0.924 & 0.969 & 0.783 \\[2pt]
& \cellcolor[rgb]{0.92,0.92,0.92} & \cellcolor[rgb]{0.92,0.92,0.92}\checkmark & \cellcolor[rgb]{0.92,0.92,0.92} & \cellcolor[rgb]{0.92,0.92,0.92}\checkmark & \cellcolor[rgb]{0.92,0.92,0.92} & \cellcolor[rgb]{0.92,0.92,0.92} & \cellcolor[rgb]{0.92,0.92,0.92}0.252 & \cellcolor[rgb]{0.92,0.92,0.92}-0.379 & \cellcolor[rgb]{0.92,0.92,0.92}0.251 & \cellcolor[rgb]{0.92,0.92,0.92}7.290 & \cellcolor[rgb]{0.92,0.92,0.92}1.665 & \cellcolor[rgb]{0.92,0.92,0.92}1.677 & \cellcolor[rgb]{0.92,0.92,0.92}1.454 & \cellcolor[rgb]{0.92,0.92,0.92}1.055 & \cellcolor[rgb]{0.92,0.92,0.92}0.924 & \cellcolor[rgb]{0.92,0.92,0.92}0.924 & \cellcolor[rgb]{0.92,0.92,0.92}0.963 & \cellcolor[rgb]{0.92,0.92,0.92}0.783 \\[2pt]
& \checkmark &  &  &  & \checkmark &  & 0.204 & -0.393 & 3.974 & 8.687 & 1.709 & 1.725 & 1.363 & 1.061 & 0.938 & 0.938 & 0.930 & 0.748 \\[2pt]
& \cellcolor[rgb]{0.92,0.92,0.92} & \cellcolor[rgb]{0.92,0.92,0.92}\checkmark & \cellcolor[rgb]{0.92,0.92,0.92} & \cellcolor[rgb]{0.92,0.92,0.92} & \cellcolor[rgb]{0.92,0.92,0.92}\checkmark & \cellcolor[rgb]{0.92,0.92,0.92} & \cellcolor[rgb]{0.92,0.92,0.92}0.168 & \cellcolor[rgb]{0.92,0.92,0.92}-0.466 & \cellcolor[rgb]{0.92,0.92,0.92}3.900 & \cellcolor[rgb]{0.92,0.92,0.92}8.643 & \cellcolor[rgb]{0.92,0.92,0.92}1.726 & \cellcolor[rgb]{0.92,0.92,0.92}1.738 & \cellcolor[rgb]{0.92,0.92,0.92}1.379 & \cellcolor[rgb]{0.92,0.92,0.92}1.067 & \cellcolor[rgb]{0.92,0.92,0.92}0.937 & \cellcolor[rgb]{0.92,0.92,0.92}0.938 & \cellcolor[rgb]{0.92,0.92,0.92}0.927 & \cellcolor[rgb]{0.92,0.92,0.92}0.748 \\[2pt]
& \checkmark &  &  &  &  & \checkmark & 0.288 & -0.313 & 0.318 & 7.299 & 1.648 & 1.663 & 1.644 & 1.052 & 0.935 & 0.936 & 0.976 & 0.805 \\[2pt]
& \cellcolor[rgb]{0.92,0.92,0.92} & \cellcolor[rgb]{0.92,0.92,0.92}\checkmark & \cellcolor[rgb]{0.92,0.92,0.92} & \cellcolor[rgb]{0.92,0.92,0.92} & \cellcolor[rgb]{0.92,0.92,0.92} & \cellcolor[rgb]{0.92,0.92,0.92}\checkmark & \cellcolor[rgb]{0.92,0.92,0.92}0.252 & \cellcolor[rgb]{0.92,0.92,0.92}-0.387 & \cellcolor[rgb]{0.92,0.92,0.92}0.275 & \cellcolor[rgb]{0.92,0.92,0.92}7.228 & \cellcolor[rgb]{0.92,0.92,0.92}1.667 & \cellcolor[rgb]{0.92,0.92,0.92}1.677 & \cellcolor[rgb]{0.92,0.92,0.92}1.456 & \cellcolor[rgb]{0.92,0.92,0.92}1.061 & \cellcolor[rgb]{0.92,0.92,0.92}0.936 & \cellcolor[rgb]{0.92,0.92,0.92}0.935 & \cellcolor[rgb]{0.92,0.92,0.92}0.969 & \cellcolor[rgb]{0.92,0.92,0.92}0.805 \\
\bottomrule
\end{tabularx}
\begin{minipage}{\textwidth}
\footnotesize \textbf{Note:} 
\begin{itemize}
    \item \textbf{W.PS} represents the survey-weighted propensity score model, where survey weights are directly used as regression weights in the propensity score estimation.
    \item \textbf{C.PS} refers to the covariate-adjusted propensity score model, where survey weights are used as an additional covariate in the propensity score estimation.
    \item \textbf{PSW} represents the propensity score weighting estimator without augmentation.
    \item \textbf{MOM} stands for the Moment Estimator.
    \item \textbf{CVR} stands for the Clever Covariate Estimator.
    \item \textbf{WET} represents the Weighted Regression Estimator.
    \item The four categories under \textbf{Relative Bias}, \textbf{Relative Efficiency}, and \textbf{Coverage} represent combinations of model specification correctness:
    \begin{itemize}
        \item \textbf{Cor\textbar Cor}: Both the propensity score and outcome models are correctly specified.
        \item \textbf{Mis\textbar Cor}: The outcome model is correctly specified, while the propensity score model is misspecified.
        \item \textbf{Cor\textbar Mis}: The propensity score model is correctly specified, while the outcome model is misspecified.
        \item \textbf{Mis\textbar Mis}: Both the propensity score and outcome models are misspecified.
    \end{itemize}
    \item The \textbf{\textendash} symbol indicates scenarios where the estimator does not involve the outcome model, hence no results are reported.
\end{itemize}
\end{minipage}
\end{sidewaystable}

\begin{sidewaystable}[htbp]
\centering
\caption{Relative bias (\%), relative efficiency, and coverage of the 95\% confidence intervals for the estimators for continuous outcomes under poor overlap across 5000 simulations.} \label{PoorOverlapSimulationResults}
\label{ExtractedPoorOverlapSimulationResults}
\scriptsize
\renewcommand{\arraystretch}{0.8}
\setlength{\tabcolsep}{0.5pt}
\begin{tabularx}{\textwidth}{@{}>{\centering\arraybackslash}p{1.5cm} *{6}{>{\centering\arraybackslash}p{1.2cm}} *{12}{>{\centering\arraybackslash}p{1.2cm}} @{}}
\toprule
\multirow{2}{*}{Estimand} & \multicolumn{6}{c}{Specification} & \multicolumn{4}{c}{Relative Bias(\%)} & \multicolumn{4}{c}{Relative Efficiency} & \multicolumn{4}{c}{Coverage} \\
\cmidrule(lr){2-7} \cmidrule(lr){8-11} \cmidrule(lr){12-15} \cmidrule(lr){16-19}
& W.PS & C.PS & PSW & MOM & CVR & WEI & Cor\textbar Cor & Mis\textbar Cor & Cor\textbar Mis & Mis\textbar Mis & Cor\textbar Cor & Mis\textbar Cor & Cor\textbar Mis & Mis\textbar Mis & Cor\textbar Cor & Mis\textbar Cor & Cor\textbar Mis & Mis\textbar Mis \\
\midrule
\multirow{8}{*}{PATE}
& \checkmark &  & \checkmark &  &  &  & 8.848 & 36.838 & \textendash & \textendash & 1.000 & 1.073 & \textendash & \textendash & 0.761 & 0.505 & \textendash & \textendash \\[2pt]
& \cellcolor[rgb]{0.92,0.92,0.92} & \cellcolor[rgb]{0.92,0.92,0.92}\checkmark & \cellcolor[rgb]{0.92,0.92,0.92}\checkmark & \cellcolor[rgb]{0.92,0.92,0.92} & \cellcolor[rgb]{0.92,0.92,0.92} & \cellcolor[rgb]{0.92,0.92,0.92} & \cellcolor[rgb]{0.92,0.92,0.92}9.192 & \cellcolor[rgb]{0.92,0.92,0.92}37.923 & \cellcolor[rgb]{0.92,0.92,0.92}\textendash & \cellcolor[rgb]{0.92,0.92,0.92}\textendash & \cellcolor[rgb]{0.92,0.92,0.92}0.984 & \cellcolor[rgb]{0.92,0.92,0.92}1.044 & \cellcolor[rgb]{0.92,0.92,0.92}\textendash & \cellcolor[rgb]{0.92,0.92,0.92}\textendash & \cellcolor[rgb]{0.92,0.92,0.92}0.761 & \cellcolor[rgb]{0.92,0.92,0.92}0.504 & \cellcolor[rgb]{0.92,0.92,0.92}\textendash & \cellcolor[rgb]{0.92,0.92,0.92}\textendash \\[2pt]
& \checkmark &  &  & \checkmark &  &  & 0.727 & 1.000 & 2.573 & 33.376 & 4.182 & 4.402 & 1.463 & 0.797 & 0.902 & 0.904 & 0.924 & 0.595 \\[2pt]
& \cellcolor[rgb]{0.92,0.92,0.92} & \cellcolor[rgb]{0.92,0.92,0.92}\checkmark & \cellcolor[rgb]{0.92,0.92,0.92} & \cellcolor[rgb]{0.92,0.92,0.92}\checkmark & \cellcolor[rgb]{0.92,0.92,0.92} & \cellcolor[rgb]{0.92,0.92,0.92} & \cellcolor[rgb]{0.92,0.92,0.92}0.744 & \cellcolor[rgb]{0.92,0.92,0.92}1.016 & \cellcolor[rgb]{0.92,0.92,0.92}2.811 & \cellcolor[rgb]{0.92,0.92,0.92}33.956 & \cellcolor[rgb]{0.92,0.92,0.92}4.211 & \cellcolor[rgb]{0.92,0.92,0.92}4.384 & \cellcolor[rgb]{0.92,0.92,0.92}1.450 & \cellcolor[rgb]{0.92,0.92,0.92}0.777 & \cellcolor[rgb]{0.92,0.92,0.92}0.903 & \cellcolor[rgb]{0.92,0.92,0.92}0.904 & \cellcolor[rgb]{0.92,0.92,0.92}0.923 & \cellcolor[rgb]{0.92,0.92,0.92}0.592 \\[2pt]
& \checkmark &  &  &  & \checkmark &  & 0.351 & 0.358 & 24.972 & 22.415 & 23.756 & 23.663 & 12.343 & 13.261 & 0.800 & 0.803 & 0.330 & 0.365 \\[2pt]
& \cellcolor[rgb]{0.92,0.92,0.92} & \cellcolor[rgb]{0.92,0.92,0.92}\checkmark & \cellcolor[rgb]{0.92,0.92,0.92} & \cellcolor[rgb]{0.92,0.92,0.92} & \cellcolor[rgb]{0.92,0.92,0.92}\checkmark & \cellcolor[rgb]{0.92,0.92,0.92} & \cellcolor[rgb]{0.92,0.92,0.92}0.340 & \cellcolor[rgb]{0.92,0.92,0.92}0.334 & \cellcolor[rgb]{0.92,0.92,0.92}24.947 & \cellcolor[rgb]{0.92,0.92,0.92}22.298 & \cellcolor[rgb]{0.92,0.92,0.92}23.733 & \cellcolor[rgb]{0.92,0.92,0.92}23.656 & \cellcolor[rgb]{0.92,0.92,0.92}12.257 & \cellcolor[rgb]{0.92,0.92,0.92}13.270 & \cellcolor[rgb]{0.92,0.92,0.92}0.788 & \cellcolor[rgb]{0.92,0.92,0.92}0.790 & \cellcolor[rgb]{0.92,0.92,0.92}0.292 & \cellcolor[rgb]{0.92,0.92,0.92}0.356 \\[2pt]
& \checkmark &  &  &  &  & \checkmark & 1.020 & 1.012 & 2.234 & 29.742 & 5.760 & 5.818 & 3.657 & 3.223 & 0.859 & 0.862 & 0.882 & 0.543 \\[2pt]
& \cellcolor[rgb]{0.92,0.92,0.92} & \cellcolor[rgb]{0.92,0.92,0.92}\checkmark & \cellcolor[rgb]{0.92,0.92,0.92} & \cellcolor[rgb]{0.92,0.92,0.92} & \cellcolor[rgb]{0.92,0.92,0.92} & \cellcolor[rgb]{0.92,0.92,0.92}\checkmark & \cellcolor[rgb]{0.92,0.92,0.92}1.022 & \cellcolor[rgb]{0.92,0.92,0.92}1.011 & \cellcolor[rgb]{0.92,0.92,0.92}2.308 & \cellcolor[rgb]{0.92,0.92,0.92}29.743 & \cellcolor[rgb]{0.92,0.92,0.92}5.761 & \cellcolor[rgb]{0.92,0.92,0.92}5.806 & \cellcolor[rgb]{0.92,0.92,0.92}3.592 & \cellcolor[rgb]{0.92,0.92,0.92}3.226 & \cellcolor[rgb]{0.92,0.92,0.92}0.858 & \cellcolor[rgb]{0.92,0.92,0.92}0.861 & \cellcolor[rgb]{0.92,0.92,0.92}0.873 & \cellcolor[rgb]{0.92,0.92,0.92}0.542 \\
\midrule
\multirow{8}{*}{PATT}
& \checkmark &  & \checkmark &  &  &  & 0.774 & 17.438 & \textendash & \textendash & 2.299 & 2.291 & \textendash & \textendash & 0.793 & 0.517 & \textendash & \textendash \\[2pt]
& \cellcolor[rgb]{0.92,0.92,0.92} & \cellcolor[rgb]{0.92,0.92,0.92}\checkmark & \cellcolor[rgb]{0.92,0.92,0.92}\checkmark & \cellcolor[rgb]{0.92,0.92,0.92} & \cellcolor[rgb]{0.92,0.92,0.92} & \cellcolor[rgb]{0.92,0.92,0.92} & \cellcolor[rgb]{0.92,0.92,0.92}1.055 & \cellcolor[rgb]{0.92,0.92,0.92}18.092 & \cellcolor[rgb]{0.92,0.92,0.92}\textendash & \cellcolor[rgb]{0.92,0.92,0.92}\textendash & \cellcolor[rgb]{0.92,0.92,0.92}2.190 & \cellcolor[rgb]{0.92,0.92,0.92}2.173 & \cellcolor[rgb]{0.92,0.92,0.92}\textendash & \cellcolor[rgb]{0.92,0.92,0.92}\textendash & \cellcolor[rgb]{0.92,0.92,0.92}0.780 & \cellcolor[rgb]{0.92,0.92,0.92}0.504 & \cellcolor[rgb]{0.92,0.92,0.92}\textendash & \cellcolor[rgb]{0.92,0.92,0.92}\textendash \\[2pt]
& \checkmark &  &  & \checkmark &  &  & -0.555 & -2.739 & 0.167 & 16.538 & 6.976 & 7.403 & 4.661 & 4.949 & 0.578 & 0.559 & 0.635 & 0.275 \\[2pt]
& \cellcolor[rgb]{0.92,0.92,0.92} & \cellcolor[rgb]{0.92,0.92,0.92}\checkmark & \cellcolor[rgb]{0.92,0.92,0.92} & \cellcolor[rgb]{0.92,0.92,0.92}\checkmark & \cellcolor[rgb]{0.92,0.92,0.92} & \cellcolor[rgb]{0.92,0.92,0.92} & \cellcolor[rgb]{0.92,0.92,0.92}-0.567 & \cellcolor[rgb]{0.92,0.92,0.92}-2.851 & \cellcolor[rgb]{0.92,0.92,0.92}0.218 & \cellcolor[rgb]{0.92,0.92,0.92}16.417 & \cellcolor[rgb]{0.92,0.92,0.92}7.002 & \cellcolor[rgb]{0.92,0.92,0.92}7.423 & \cellcolor[rgb]{0.92,0.92,0.92}4.542 & \cellcolor[rgb]{0.92,0.92,0.92}4.905 & \cellcolor[rgb]{0.92,0.92,0.92}0.580 & \cellcolor[rgb]{0.92,0.92,0.92}0.556 & \cellcolor[rgb]{0.92,0.92,0.92}0.632 & \cellcolor[rgb]{0.92,0.92,0.92}0.282 \\[2pt]
& \checkmark &  &  &  & \checkmark &  & 0.022 & -2.186 & 13.213 & 13.100 & 19.046 & 19.401 & 11.532 & 11.655 & 0.938 & 0.928 & 0.497 & 0.602 \\[2pt]
& \cellcolor[rgb]{0.92,0.92,0.92} & \cellcolor[rgb]{0.92,0.92,0.92}\checkmark & \cellcolor[rgb]{0.92,0.92,0.92} & \cellcolor[rgb]{0.92,0.92,0.92} & \cellcolor[rgb]{0.92,0.92,0.92}\checkmark & \cellcolor[rgb]{0.92,0.92,0.92} & \cellcolor[rgb]{0.92,0.92,0.92}-0.007 & \cellcolor[rgb]{0.92,0.92,0.92}-2.320 & \cellcolor[rgb]{0.92,0.92,0.92}13.119 & \cellcolor[rgb]{0.92,0.92,0.92}13.032 & \cellcolor[rgb]{0.92,0.92,0.92}19.097 & \cellcolor[rgb]{0.92,0.92,0.92}19.441 & \cellcolor[rgb]{0.92,0.92,0.92}11.328 & \cellcolor[rgb]{0.92,0.92,0.92}11.513 & \cellcolor[rgb]{0.92,0.92,0.92}0.937 & \cellcolor[rgb]{0.92,0.92,0.92}0.923 & \cellcolor[rgb]{0.92,0.92,0.92}0.523 & \cellcolor[rgb]{0.92,0.92,0.92}0.599 \\[2pt]
& \checkmark &  &  &  &  & \checkmark & -0.701 & -2.886 & 1.059 & 16.278 & 9.262 & 9.386 & 6.606 & 6.052 & 0.886 & 0.874 & 0.888 & 0.575 \\[2pt]
& \cellcolor[rgb]{0.92,0.92,0.92} & \cellcolor[rgb]{0.92,0.92,0.92}\checkmark & \cellcolor[rgb]{0.92,0.92,0.92} & \cellcolor[rgb]{0.92,0.92,0.92} & \cellcolor[rgb]{0.92,0.92,0.92} & \cellcolor[rgb]{0.92,0.92,0.92}\checkmark & \cellcolor[rgb]{0.92,0.92,0.92}-0.721 & \cellcolor[rgb]{0.92,0.92,0.92}-3.011 & \cellcolor[rgb]{0.92,0.92,0.92}1.084 & \cellcolor[rgb]{0.92,0.92,0.92}16.148 & \cellcolor[rgb]{0.92,0.92,0.92}9.321 & \cellcolor[rgb]{0.92,0.92,0.92}9.442 & \cellcolor[rgb]{0.92,0.92,0.92}6.431 & \cellcolor[rgb]{0.92,0.92,0.92}6.118 & \cellcolor[rgb]{0.92,0.92,0.92}0.885 & \cellcolor[rgb]{0.92,0.92,0.92}0.870 & \cellcolor[rgb]{0.92,0.92,0.92}0.888 & \cellcolor[rgb]{0.92,0.92,0.92}0.576 \\
\midrule
\multirow{8}{*}{PATO}
& \checkmark &  & \checkmark &  &  &  & 0.492 & 23.623 & \textendash & \textendash & 24.336 & 15.337 & \textendash & \textendash & 0.938 & 0.157 & \textendash & \textendash \\[2pt]
& \cellcolor[rgb]{0.92,0.92,0.92} & \cellcolor[rgb]{0.92,0.92,0.92}\checkmark & \cellcolor[rgb]{0.92,0.92,0.92}\checkmark & \cellcolor[rgb]{0.92,0.92,0.92} & \cellcolor[rgb]{0.92,0.92,0.92} & \cellcolor[rgb]{0.92,0.92,0.92} & \cellcolor[rgb]{0.92,0.92,0.92}0.544 & \cellcolor[rgb]{0.92,0.92,0.92}24.283 & \cellcolor[rgb]{0.92,0.92,0.92}\textendash & \cellcolor[rgb]{0.92,0.92,0.92}\textendash & \cellcolor[rgb]{0.92,0.92,0.92}15.168 & \cellcolor[rgb]{0.92,0.92,0.92}11.695 & \cellcolor[rgb]{0.92,0.92,0.92}\textendash & \cellcolor[rgb]{0.92,0.92,0.92}\textendash & \cellcolor[rgb]{0.92,0.92,0.92}0.892 & \cellcolor[rgb]{0.92,0.92,0.92}0.191 & \cellcolor[rgb]{0.92,0.92,0.92}\textendash & \cellcolor[rgb]{0.92,0.92,0.92}\textendash \\[2pt]
& \checkmark &  &  & \checkmark &  &  & 0.423 & -0.454 & 0.395 & 23.793 & 25.027 & 25.254 & 24.745 & 15.280 & 0.912 & 0.911 & 0.973 & 0.174 \\[2pt]
& \cellcolor[rgb]{0.92,0.92,0.92} & \cellcolor[rgb]{0.92,0.92,0.92}\checkmark & \cellcolor[rgb]{0.92,0.92,0.92} & \cellcolor[rgb]{0.92,0.92,0.92}\checkmark & \cellcolor[rgb]{0.92,0.92,0.92} & \cellcolor[rgb]{0.92,0.92,0.92} & \cellcolor[rgb]{0.92,0.92,0.92}0.402 & \cellcolor[rgb]{0.92,0.92,0.92}-0.565 & \cellcolor[rgb]{0.92,0.92,0.92}0.391 & \cellcolor[rgb]{0.92,0.92,0.92}23.794 & \cellcolor[rgb]{0.92,0.92,0.92}25.105 & \cellcolor[rgb]{0.92,0.92,0.92}25.287 & \cellcolor[rgb]{0.92,0.92,0.92}21.835 & \cellcolor[rgb]{0.92,0.92,0.92}15.342 & \cellcolor[rgb]{0.92,0.92,0.92}0.912 & \cellcolor[rgb]{0.92,0.92,0.92}0.912 & \cellcolor[rgb]{0.92,0.92,0.92}0.962 & \cellcolor[rgb]{0.92,0.92,0.92}0.175 \\[2pt]
& \checkmark &  &  &  & \checkmark &  & 0.364 & -0.510 & 18.058 & 21.367 & 31.746 & 31.867 & 20.604 & 19.124 & 0.938 & 0.939 & 0.347 & 0.216 \\[2pt]
& \cellcolor[rgb]{0.92,0.92,0.92} & \cellcolor[rgb]{0.92,0.92,0.92}\checkmark & \cellcolor[rgb]{0.92,0.92,0.92} & \cellcolor[rgb]{0.92,0.92,0.92} & \cellcolor[rgb]{0.92,0.92,0.92}\checkmark & \cellcolor[rgb]{0.92,0.92,0.92} & \cellcolor[rgb]{0.92,0.92,0.92}0.344 & \cellcolor[rgb]{0.92,0.92,0.92}-0.621 & \cellcolor[rgb]{0.92,0.92,0.92}17.989 & \cellcolor[rgb]{0.92,0.92,0.92}21.286 & \cellcolor[rgb]{0.92,0.92,0.92}31.862 & \cellcolor[rgb]{0.92,0.92,0.92}31.918 & \cellcolor[rgb]{0.92,0.92,0.92}20.692 & \cellcolor[rgb]{0.92,0.92,0.92}19.221 & \cellcolor[rgb]{0.92,0.92,0.92}0.938 & \cellcolor[rgb]{0.92,0.92,0.92}0.938 & \cellcolor[rgb]{0.92,0.92,0.92}0.346 & \cellcolor[rgb]{0.92,0.92,0.92}0.218 \\[2pt]
& \checkmark &  &  &  &  & \checkmark & 0.423 & -0.474 & 0.482 & 23.638 & 25.049 & 24.949 & 24.893 & 15.469 & 0.942 & 0.941 & 0.983 & 0.220 \\[2pt]
& \cellcolor[rgb]{0.92,0.92,0.92} & \cellcolor[rgb]{0.92,0.92,0.92}\checkmark & \cellcolor[rgb]{0.92,0.92,0.92} & \cellcolor[rgb]{0.92,0.92,0.92} & \cellcolor[rgb]{0.92,0.92,0.92} & \cellcolor[rgb]{0.92,0.92,0.92}\checkmark & \cellcolor[rgb]{0.92,0.92,0.92}0.405 & \cellcolor[rgb]{0.92,0.92,0.92}-0.585 & \cellcolor[rgb]{0.92,0.92,0.92}0.511 & \cellcolor[rgb]{0.92,0.92,0.92}23.505 & \cellcolor[rgb]{0.92,0.92,0.92}25.162 & \cellcolor[rgb]{0.92,0.92,0.92}25.004 & \cellcolor[rgb]{0.92,0.92,0.92}22.065 & \cellcolor[rgb]{0.92,0.92,0.92}15.619 & \cellcolor[rgb]{0.92,0.92,0.92}0.942 & \cellcolor[rgb]{0.92,0.92,0.92}0.941 & \cellcolor[rgb]{0.92,0.92,0.92}0.976 & \cellcolor[rgb]{0.92,0.92,0.92}0.224 \\
\bottomrule
\end{tabularx}
\begin{minipage}{\textwidth}
\footnotesize \textbf{Note:} 
\begin{itemize}
    \item \textbf{W.PS} represents the survey-weighted propensity score model, where survey weights are directly used as regression weights in the propensity score estimation.
    \item \textbf{C.PS} refers to the covariate-adjusted propensity score model, where survey weights are used as an additional covariate in the propensity score estimation.
    \item \textbf{PSW} represents the propensity score weighting estimator without augmentation.
    \item \textbf{MOM} stands for the Moment Estimator.
    \item \textbf{CVR} stands for the Clever Covariate Estimator.
    \item \textbf{WET} represents the Weighted Regression Estimator.
    \item The four categories under \textbf{Relative Bias}, \textbf{Relative Efficiency}, and \textbf{Coverage} represent combinations of model specification correctness:
    \begin{itemize}
        \item \textbf{Cor\textbar Cor}: Both the propensity score and outcome models are correctly specified.
        \item \textbf{Mis\textbar Cor}: The outcome model is correctly specified, while the propensity score model is misspecified.
        \item \textbf{Cor\textbar Mis}: The propensity score model is correctly specified, while the outcome model is misspecified.
        \item \textbf{Mis\textbar Mis}: Both the propensity score and outcome models are misspecified.
    \end{itemize}
    \item The \textbf{\textendash} symbol indicates scenarios where the estimator does not involve the outcome model, hence no results are reported.
\end{itemize}
\end{minipage}
\end{sidewaystable}

\subsubsection{How should survey weights be incorporated into the propensity score weighting estimator? } \label{sec4.2.1}

Focusing first on the results from the weighting estimator alone (PSW), our simulation results support our theoretical conclusion that incorporating survey weights in both the propensity score estimation and final standardization step leads to unbiased estimation of the population-level treatment effect. 
Under good overlap (Table \ref{GoodOverlapSimulationResults}), using survey-weighted propensity score models (W.PS) consistently produces lower bias, higher relative efficiency, and coverage closer to the nominal 95\% compared to using survey weights as an additional covariate (C.PS). This pattern is particularly evident for PATO, where W.PS shows satisfactory performance across all specification scenarios. Overall, the results indicate that W.PS excels when the propensity score model is correct and continues to outperform C.PS even under a misspecified propensity score model (the "Mis\textbar Cor" columns).

Under poor overlap (Table \ref{PoorOverlapSimulationResults}), the advantage of W.PS over C.PS becomes more pronounced. While C.PS suffers from inflated bias and poor coverage, particularly for estimating PATE and PATT, W.PS maintains lower bias, higher efficiency, and closer-to-nominal coverage across all estimands. It is worth noting that inference for PATO is more robust to poor overlap, as PSW with W.PS achieves the lowest biases and highest relative efficiency among all PSW methods. Moreover, PSW with W.PS for PATO maintains near-nominal coverage at 93.8\%, while C.PS drops below 90\%. 
Even when the propensity score model is misspecified and interaction terms are ignored (“Mis\textbar Cor” columns in Tables \ref{GoodOverlapSimulationResults} and \ref{PoorOverlapSimulationResults}), W.PS consistently outperforms C.PS in terms of bias and coverage. 

Additional empirical evidence from additional simulations (Web Tables 9 and 10), where sampling depends on treatment assignment, lead to the same findings. In these extended scenarios, W.PS achieves near-zero bias, while the relative bias of C.PS deteriorates to 25\%–54\% even under good overlap (Web Table 9).
Extreme survey weights, likely when sampling depends on treatment, exacerbate model extrapolation and misspecification issues for C.PS. Inference for PATO demonstrates outstanding robustness in the alternative sampling mechanism, especially under poor overlap (Web Table 10). Importantly, PSW with W.PS for PATO achieves the smallest bias (0.5\%) and the highest efficiency---nearly 30 times the reference method (PSW with W.PS for PATE)
---and is the only PSW estimator maintaining nominal coverage (94.4\%) under this challenging scenario.
These findings align with insights from Ridgeway et al.,\cite{Ridgeway2015} which highlight the benefits of survey-weighted propensity scores across various sampling mechanisms, though their previous investigation was restricted to PATT. Our results cover additional estimands (PATE, PATT, and PATO) different overlap and misspecification scenarios.

The above findings also apply to augmented estimators with some nuances. In our main simulation scenarios (Tables \ref{GoodOverlapSimulationResults} and \ref{PoorOverlapSimulationResults}), augmented estimators with W.PS or C.PS perform similarly; this indicates that augmentation provides some protective effect and confers a degree of robustness regardless of how the propensity scores are estimated. However, under treatment-dependent sampling (Web Tables 9 and 10), 
the differences become substantial, and W.PS wins over C.PS in both bias and coverage even with augmentation.
Overall, our empirical results highlight W.PS as the preferred method for incorporating survey weights into propensity score estimation, regardless of whether an outcome regression component is used to construct the final causal effect estimator.

\subsubsection{Is there a global winner among the augmented estimators when all models are correctly specified?} \label{sec4.2.2}

Our simulation results, when both the propensity score and outcome models are correctly specified, indicate that all three augmented estimators consistently outperform PSW alone. While CVR demonstrate high efficiency under good overlap (Table \ref{GoodOverlapSimulationResults}), it can be sensitive to different sampling mechanisms and overlap scenarios. MOM offers consistent robustness but provides less bias control and coverage stability compared to WET across all estimands. Consequently, WET with W.PS emerges as the most reliable and efficient estimator in finite samples, maintaining near-nominal coverage, lower bias, particularly under poor overlap (Table \ref{PoorOverlapSimulationResults}) and treatment-dependent sampling (Web Table 9 and 10).

To be more explicit, we first focus on the "Cor\textbar Cor" columns under good overlap (Table \ref{GoodOverlapSimulationResults}), W.PS-based MOM, CVR, and WET outperform PSW across all estimands (PATE, PATT, and PATO).
The CVR estimator achieves significant efficiency gains, notably for PATT and PATO. 
Meanwhile, 
WET consistently performs slightly better than MOM, maintaining around 93\% coverage and biases below 2\% across most rows.

Under poor overlap ("Cor\textbar Cor" columns in Table \ref{PoorOverlapSimulationResults})
PSW alone produces noticeable biases (e.g., 8.848\% for PATE) and reduced coverage (76.1\%). This highlights the well-known vulnerability of simple weighting methods under poor overlap.\cite{Li2019} 
In contrast, MOM, CVR, and WET demonstrate lower biases and improved efficiency and coverage.
For example, CVR with W.PS achieves nearly twenty-fold efficiency gains for PATE, reaching a relative efficiency of 23.8.
Similarly, CVR boosts efficiency for PATT and PATO while maintaining better control over coverage. However, comparing across the augmented estimators, CVR and MOM show greater sensitivity under poor overlap. For instance, CVR achieves only 80\% coverage for PATE, and MOM drops to 57.8\% coverage for PATT. In comparison, WET consistently maintains coverage near or above 90\%. Among these, WET with W.PS stands out, achieving the highest nominal coverage for PATO at 94.2\% and outperforming other estimators.

When sampling probabilities depend on treatment assignment (Web Tables 9 and 10), the advantage of WET becomes more evident. Under good overlap scenarios (Web Table 9), WET with W.PS achieves minimal biases and relative efficiency exceeding 1.5 for PATE and PATT while maintaining nominal coverage around 94\%. By contrast, CVR suffers from larger bias and under-coverage, particularly for PATT. 
Similar patterns are found under poor overlap (Web Table 10).

Taken together, WET with W.PS excels under complex sampling mechanisms. The weighted regression framework effectively minimizes bias and maintains stable coverage, even under challenging scenarios with poor overlap. While MOM provides consistent robustness and CVR achieves high efficiency under ideal conditions as in Table \ref{GoodOverlapSimulationResults}, both tend to be unstable when survey weights are increasingly heterogeneous. Thus, WET with W.PS stands out as a robust choice in our empirical investigation, particularly when the sampling mechanism depends on treatment and generates highly variable survey weights.

\subsubsection{Under misspecification, which methods are more robust and should be recommended under the complex survey setting?} \label{sec4.2.3}

When model specifications deviate from the truth, the performance of estimators varies by estimand and overlap. 
When the propensity score model is misspecified but the outcome model is correct under good overlap (“Mis\textbar Cor” columns in Table \ref{GoodOverlapSimulationResults}), all three augmented estimators achieve nominal coverage for PATE, demonstrating their doubly robust properties compared to PSW alone, which struggles to achieve even 80\% coverage.
MOM and WET control bias and coverage better than CVR.
In poor overlap scenarios (Table \ref{PoorOverlapSimulationResults}), CVR achieves smaller bias 
but suffers reduced coverage (80\%). In contrast, WET with W.PS maintains robustness, achieving 94.1\% coverage for PATO, even with misspecified propensity score models.

When the propensity score model is correct but the outcome model is misspecified (“Cor\textbar Mis” columns), CVR demonstrates notable instability. Its bias increases dramatically for PATE under good overlap (Table \ref{GoodOverlapSimulationResults}), reaching nearly 10\% compared to 0.2\% bias for MOM and WET, and its coverage drops below 50\% under poor overlap (Table \ref{PoorOverlapSimulationResults}).
This empirical finding highlights the reliance of CVR on the correct outcome model to manage the extreme values introduced by clever covariates in poor overlap scenarios. In contrast, WET consistently performs best,  achieving the optimal efficiency-coverage balance for PATO, despite the poor overlap and outcome model misspecification.

Under the most challenging scenario, where both the propensity score and outcome models are misspecified (“Mis\textbar Mis” columns), all methods experience significant biases and deteriorated coverage. Despite these challenges, WET remains the most reliable option, consistently maintaining coverage around 80–85\% under good overlap, while CVR and MOM frequently see coverage drop below 80\%. The extent of bias varies depending on the type of misspecification. In our simulation settings, where interaction terms are omitted from both models, this misspecification strongly affects propensity score estimation and outcome regression, leading to more biased estimates and poor coverage across all estimators.

Additional simulations (Web Tables 9 and 10), where sampling depends on treatment and  extreme survey weights are present, further highlight the relative advantage of WET. WET with W.PS consistently preserves its reliability across misspecified models and overlap conditions, achieving minimal bias (less than 3\%) and near-nominal coverage. CVR, by contrast, becomes highly vulnerable under outcome model misspecification, with biases escalating to beyond 70–100\% and coverage collapsing to near zero for PATT under poor overlap. MOM remains more stable than CVR but falls short of WET in terms of overall consistency.

In conclusion, while CVR demonstrates potential for high efficiency with correctly specified models under good overlap, its heightened sensitivity to model misspecification and poor overlap may limit its practical reliability.
In contrast, both MOM and WET exhibit greater robustness, with WET emerging as the most resilient choice across varying simulation settings.

\section{Two case studies}\label{sec5}

\subsection{Application to the ECLS-K dataset}

In this first data example, we present a real-world example with multi-stage sampling to demonstrate the application of the different methods for using propensity score weighting with complex survey data. The data used are from the Early Childhood Longitudinal Study, Kindergarten class 1998-1999 (ECLS-K), \cite{Tourangeau2009} a nationally representative longitudinal study that tracks early school experiences of a large cohort from kindergarten through eighth grade in the United States. The study collects comprehensive information at the child, household, and school levels, with survey weights reflecting the multi-stage probability sampling design employed to ensure representativeness. \cite{NationalCenter2011} 
The participants were selected prior to the assignment of treatment. Specifically, children were sampled at the start of the study upon their entry into kindergarten, and subsequent treatment, such as the receipt of special education services, was determined in later stages, typically based on administrative records collected during follow-up data collection waves, such as in first grade or fifth grade. 
This example reflects a prospective study design, consistent with our main simulation setting, in which the sampling process does not depend on treatment assignment and survey weights are explicitly known. This design simplifies the integration of survey weights into subsequent estimations by reducing the steps required. Corresponding estimators tailored for prospective studies are further detailed in the appendix.
The treatment status was defined as participation in special education services, as documented in school administrative records. The outcome of interest was the IRT-scaled math achievement test score from the ECLS-K dataset. A set of 39 baseline covariates covering demographic, household, and school characteristics were included to address potential confounding, as described in Keller and Tipton.\cite{Keller2016} 

\begin{figure}[!htbp]
\centering
\begin{minipage}{0.49\textwidth}
    \centering
    \includegraphics[width=\textwidth]{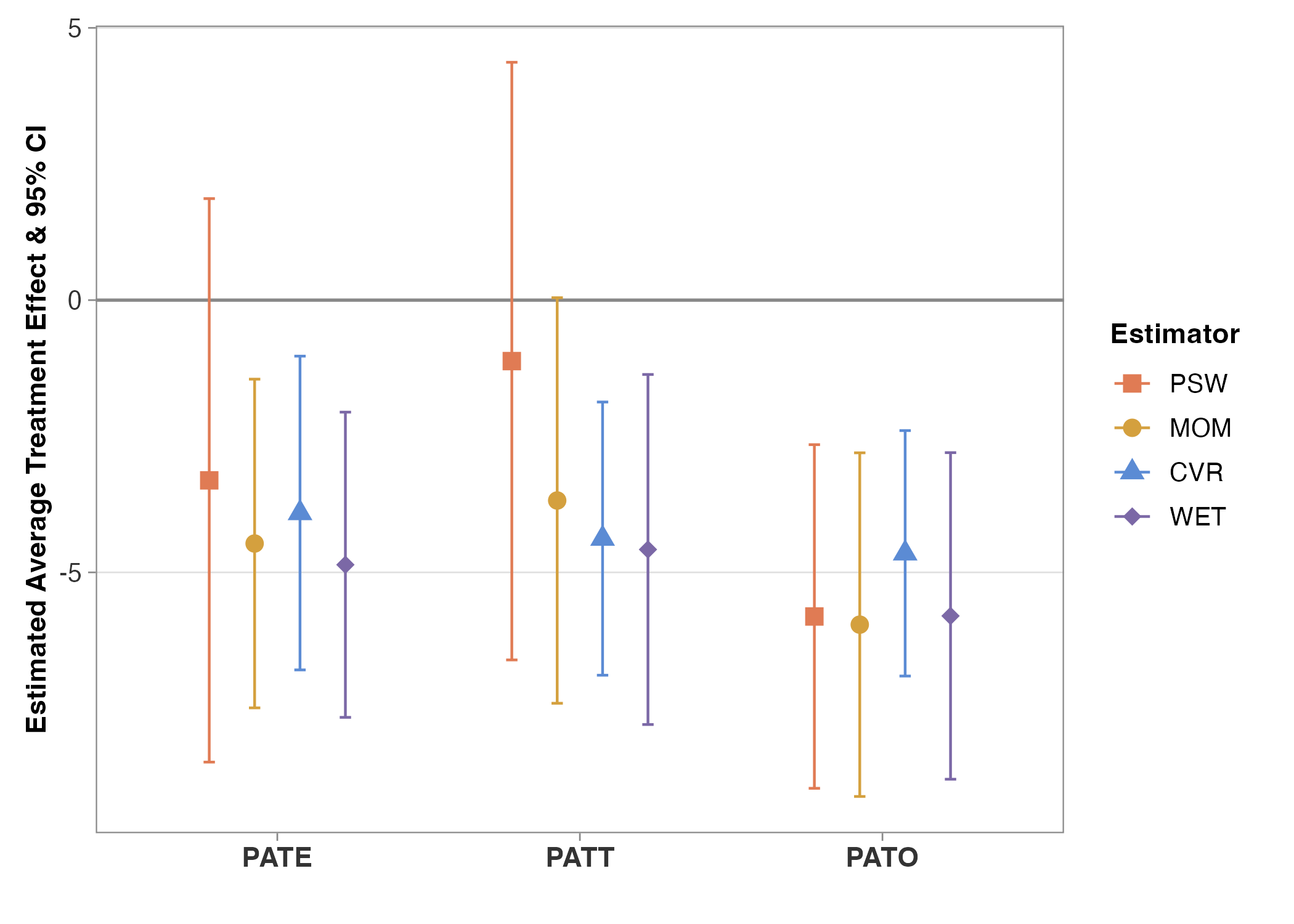}
    \subcaption{ECLS-K: Effect of special education on math achievement}
    \label{fig:forest_ecls}
\end{minipage}

\begin{minipage}{0.49\textwidth}
    \centering
    \includegraphics[width=\textwidth]{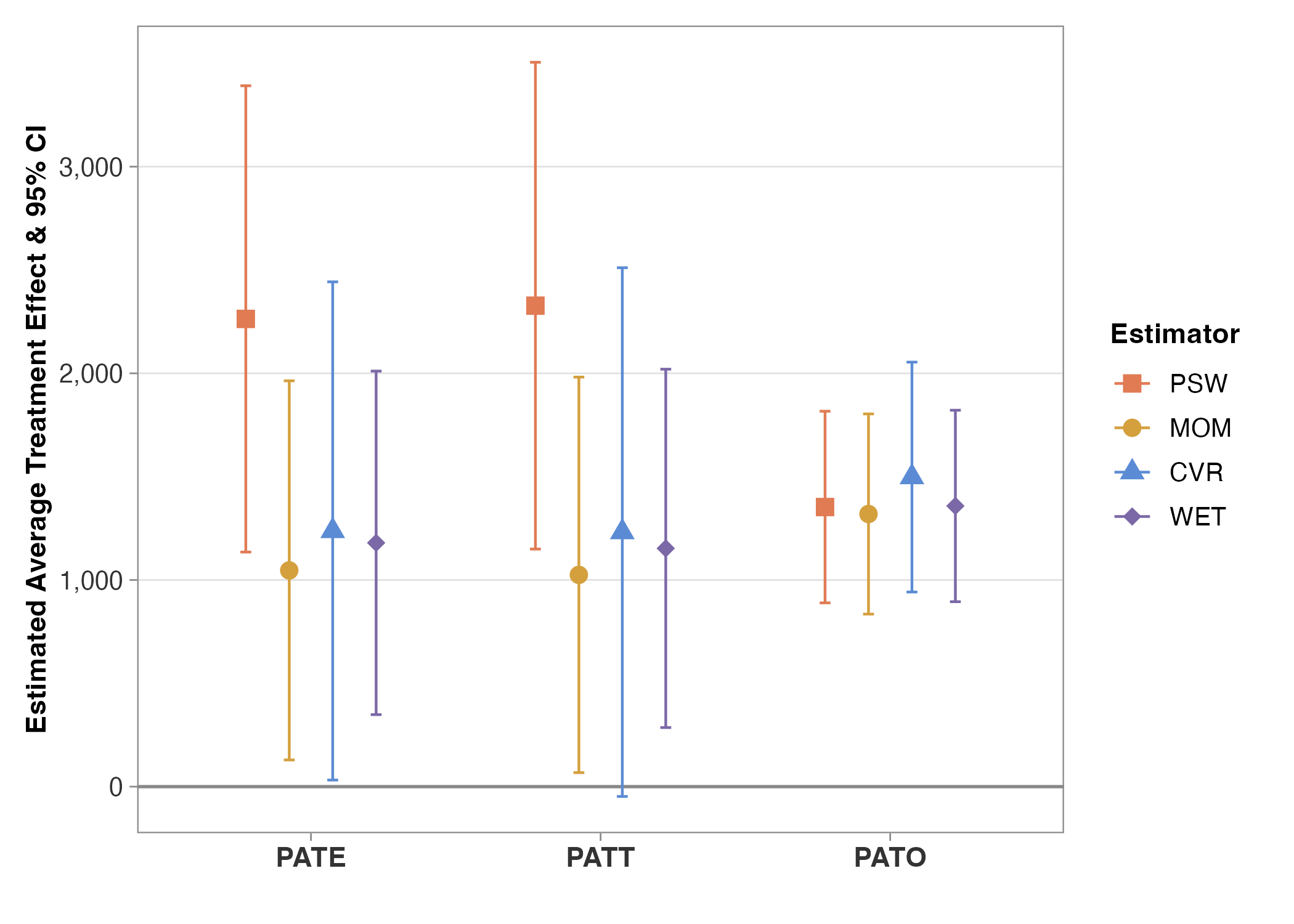}
    \subcaption{MEPS: White–Asian comparison}
    \label{fig:forest_meps_asian}
\end{minipage}
\hfill
\begin{minipage}{0.49\textwidth}
    \centering
    \includegraphics[width=\textwidth]{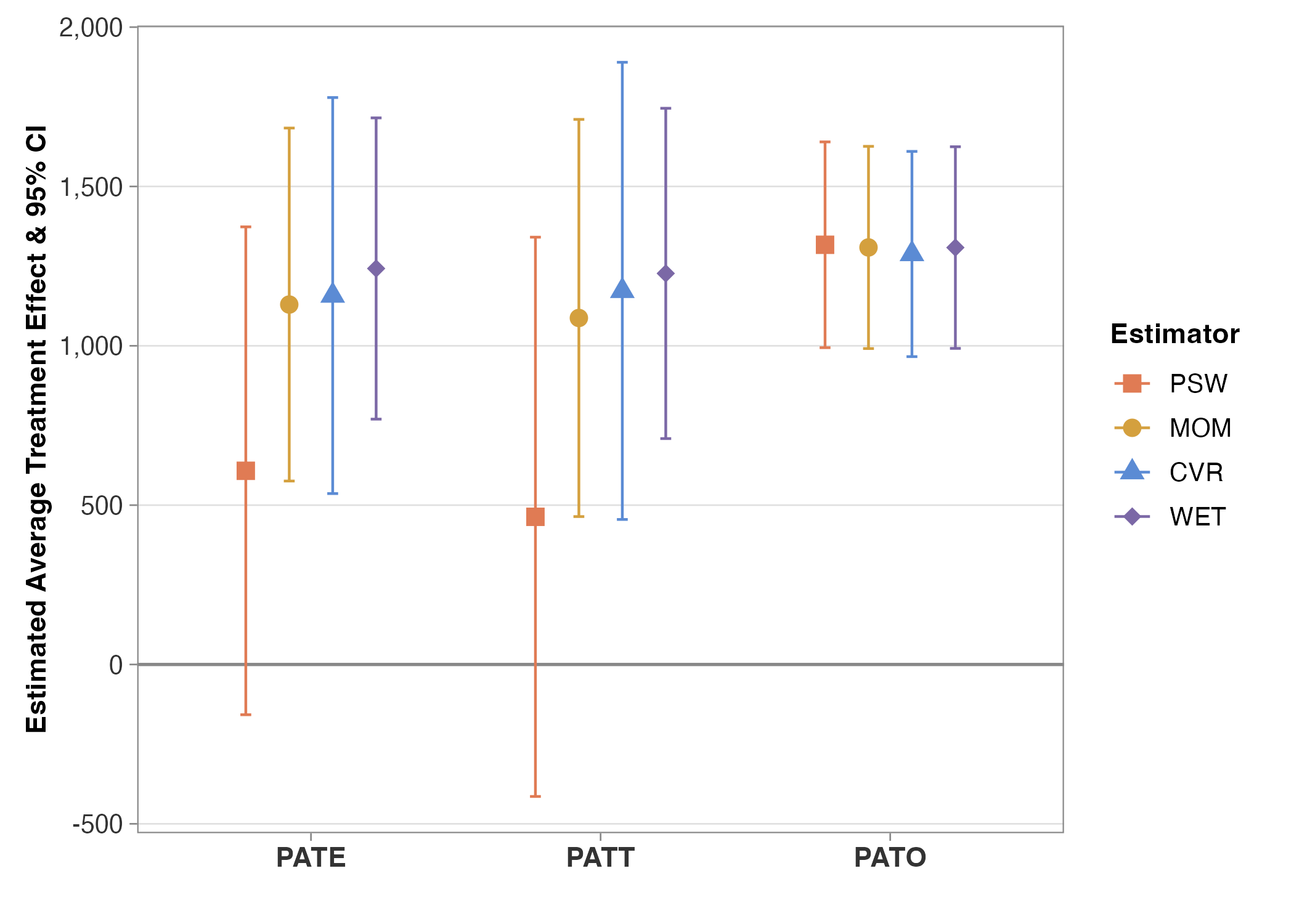}
    \subcaption{MEPS: White–Hispanic comparison}
    \label{fig:forest_meps_hispanic}
\end{minipage}
\caption{Estimated treatment effects with 95\% confidence intervals across estimators (PSW, MOM, CVR, WET) and estimands (PATE, PATT, PATO). Panel (a) shows the effect of special education services on fifth-grade math achievement in the ECLS-K study. Panels (b) and (c) show racial disparities in annual healthcare expenditures (2009 USD) in the MEPS study. The horizontal line indicates zero effect or disparity.}
\label{fig:forest_combined}
\end{figure}

Web Figure 1 
illustrates the population-level covariate balance, measured by Population-level survey-weighted Standardized Mean Differences (PSMD), which align with the weights \((\omega_{1,p}(\mathbf{X}_i), \omega_{0,p}(\mathbf{X}_i))\) used in the effect estimation and correspond to the specifications outlined in Table \ref{Table1CommonBalancingWeightsUnderSurveySettings} for each target population and estimand. 
The PSMD for a baseline covariate \( X \) is calculated as the absolute difference between the weighted means of \( X \) for the treated and control groups, scaled by the pooled weighted standard deviation \( s_p \). Specifically, PSMD is defined as \(|\bar{X}_1 - \bar{X}_0| / s_p\), where the weighted means are \(\bar{X}_1 = \left(\sum_{i=1}^{n} \omega_{1,p}(\mathbf{X}_i)  X_i  Z_i\right) / \left(\sum_{i=1}^{n} \omega_{1,p}(\mathbf{X}_i)  Z_i\right)\) for the treated group and \(\bar{X}_0 = \left(\sum_{i=1}^{n} \omega_{0,p}(\mathbf{X}_i)  X_i  (1 - Z_i)\right) / \left(\sum_{i=1}^{n} \omega_{0,p}(\mathbf{X}_i)  (1 - Z_i)\right)\) for the control group. The pooled weighted standard deviation, \( s_p \), reflects a composite measure of the variability of \( X \) within the treated and control groups, weighted according to the propensity score weights \( \omega_{1,p} \) and \( \omega_{0,p} \), and accounts for the sample size in each group. 
The figure compares the covariate balance for each population-level estimand—PATE, PATT, and PATO—using survey-weighted regression (W.PS). The visual evidence clearly demonstrates that W.PS achieves superior covariate balance, as indicated by a larger proportion of covariates achieving PSMDs below the 0.10 threshold, a conventional threshold for adequate balance.\cite{Austin2015} This is further detailed in the grey-shaded cells of Web Table 17, which shows how W.PS systematically improves covariate balance across most baseline characteristics. Specifically, W.PS achieves exact balance for PATO, as a result of Proposition \ref{ExactBalance}. We then estimate the average causal effect of elementary school special education services on fifth-grade math achievement, and the results are presented in Figure \ref{fig:forest_combined}(a), with numerical details provided in Web Table 27.

Across the three estimands targeting distinct populations, the results reveal notable differences in precision and stability. The PATE estimates the average effect across the entire student population, yielding negative effects ranging from $-3.31$ to $-4.86$ depending on the estimator, with PSW showing particularly large standard errors (SE $= 2.64$). The PATT focuses on students who actually received special education services, where most of the augmented estimators (MOM, CVR, WET) yield statistically significant estimates, while PSW ($-1.12$, SE $= 2.80$) fails to detect the effect and produces a much wider confidence interval including zero. The PATO targets students for whom the special education placement decision is most uncertain and comparative evidence is most needed. For this estimand, all estimators produce more similar point estimates (ranging from $-4.65$ to $-5.96$) with comparable or smaller standard errors. For PATO, the WET estimate indicates that among these comparable students, those receiving special education scored approximately 5.8 points lower in math. Interestingly, among the augmented estimators, WET demonstrates particularly stable performance across all three estimands. These patterns align with our simulation observations.

\subsection{Application to the 2009 Medical Expenditure Panel Survey dataset}

In our second example, we demonstrate how the proposed methods perform in a retrospective survey design where the sampling mechanism depends directly on the exposure, specifically race. 
We use data from the 2009 Medical Expenditure Panel Survey (MEPS),\cite{AHRQ2011} which historically oversamples minority groups, including Hispanics and Blacks, to improve the precision of health disparity analyses. Starting in 2006, this strategy expanded to include Asian households in response to changing demographics and analytical priorities.\cite{chowdhury2019sample} This oversampling design ensures adequate representation of minority groups but can also lead to disproportionate survey weights, particularly for subgroups with limited population overlap. 
Unlike Data Example 1 (ECLS-K), where participants were sampled before treatment status was realized, MEPS employs a retrospective design where individuals are sampled after their race is determined, leading to survey weights that can correlate with race and demographic strata. 

The 2009 MEPS public-release file includes 9830 non-Hispanic White adults (hereafter “Whites”), 4020 Blacks, 1446 Asians, and 5150 Hispanics, each assigned a survey weight reflecting their probability of selection and nonresponse adjustments.\cite{Cook2010} Here, we focus on two pairwise comparisons (White–Asian and White–Hispanic) to estimate racial disparities in total annual healthcare expenditures, measured in 2009 USD. Although “race” is not a manipulable treatment, the methodological goal parallels standard propensity score approaches to balancing covariates between two groups. Following the framework for unconfounded descriptive comparisons described in Li and Li,\cite{Li2023} we control for a set of variables reflecting clinical need, as instructed by the Institute of Medicine (IOM),\cite{McGuire2006} including body mass index (BMI), SF-12 physical and mental component summaries, health conditions, age, gender, and marital status. 
Additionally, we account for selected socioeconomic and functional limitation variables, consistent with prior applications to the MEPS data without adjusting for survey weights.\cite{Li2023}

Web Figures 2, 3, and 4 
collectively illustrate the interplay between survey weights, propensity score distributions estimated by the survey-weighted propensity score model (W.PS), and covariate balance under different weighting methods for White–Asian and White–Hispanic comparisons. In the White–Asian comparison, MEPS’s oversampling strategy leads to highly concentrated survey weights for Asians and a broader range with extreme weights for Whites, amplifying disparities in baseline covariates like BMI. These imbalances are reflected in the propensity score distributions, estimated using the survey-weighted propensity score model (W.PS), which show limited overlap between the groups, with most Whites concentrated at high scores (above 0.9) and Asians clustering near 0.8–0.9. The love plots further demonstrate that this limited overlap results in poor covariate balance under PATE and PATT, particularly under C.PS, where many covariates exceed the 0.1 PSMD threshold. Conversely, overlap weighting (PATO), especially under W.PS, achieves exact balance by focusing on comparable subpopulations, mitigating instability caused by extreme weights. In the White–Hispanic comparison, stronger overlap and more balanced survey weights yield better covariate alignment across methods, with PATO consistently outperforming other estimands in balancing covariates in the target population.

Figure \ref{fig:forest_combined}(b) and (c) display the estimated differences in annual healthcare expenditures (in 2009 USD) between Whites and Asians and between Whites and Hispanics, with numerical details provided in Web Table 28. In the White–Asian comparison, limited overlap leads to substantial variability under PSW, with PATE and PATT estimates showing large standard errors (SE $= 575.60$ and $600.88$, respectively). The augmented estimators improve precision considerably, with WET yielding a PATE estimate of \$1179.61 (SE $= 424.04$). The PATO emphasizes a naturally comparable subpopulation with similar health status that is well-represented among both Whites and Asians. For this estimand, all estimators produce more similar estimates (ranging from \$1319.22 to \$1497.99) with consistently smaller standard errors around 236 to 284. In the White–Hispanic comparison, stronger overlap yields better precision overall compared to the White–Asian comparison. However, PSW estimates for PATE (\$607.78, SE $= 390.54$) and PATT (\$463.37, SE $= 447.72$) still show notable variability, while PATO estimates cluster tightly around \$1288 to \$1317 with standard errors near 161 to 165. Among the augmented estimators, WET again achieves the smallest standard errors across both comparisons, yielding PATO estimates of \$1358.16 (SE $= 236.36$) for White–Asian and \$1308.27 (SE $= 161.41$) for White–Hispanic, matching the observations in our simulation studies.

\section{Discussion}\label{sec6}

We introduce a unified framework for incorporating survey weights into the balancing weights estimators for population-level causal inference. 
We establish the asymptotic normality of survey-weighted estimators for retrospective designs (with prospective designs detailed in the Web Appendix), and provide a robust sandwich variance estimator for weighting and augmented estimators. Simulation studies evaluate these methods under varying overlap and model misspecification conditions, and highlight weighted regression estimator as a numerically stable approach even in challenging poor overlap settings. 
These case studies highlight the strengths of survey-weighted propensity score models (W.PS) in achieving population-level covariate balance, WET in providing robust estimation, and the PATO estimand in enabling reliable inference within comparable subpopulations. 
All proposed methods are implemented in the publicly available PSweight R package.

Several lessons have emerged from our exploration. First, incorporating survey weights throughout all phases of the propensity score weighting estimator is essential for unbiased treatment effects estimation in survey observational studies. Using a survey-weighted regression model (W.PS) for propensity score estimation aligns sampled data covariate distributions with the target population, enhancing efficiency, coverage, and robustness across varying overlap scenarios. Consistent with Ridgeway et al.,\cite{Ridgeway2015} our results suggest that omitting survey weights restricts balance to within the sample,\cite{Zanutto2006, DugoffSchulerStuart2014} potentially introducing bias when generalizing to target populations. Consequently, W.PS is integral to valid population-level causal inferences. Our investigation clarifies when survey-weighted propensity scores are necessary versus beneficial. Under retrospective (informative) sampling where $e_{\text{fp}}(\mathbf{x}) \neq e_{\text{sp}}(\mathbf{x})$, Equation \eqref{eq:efp-retro-ps} shows that survey-weighted estimation is essential for recovering the population propensity score required by the balancing weights. Under prospective (non-informative) sampling where these quantities coincide, either approach is theoretically valid. However, our simulations (Tables \ref{GoodOverlapSimulationResults}-\ref{PoorOverlapSimulationResults}) demonstrate consistent practical advantages of survey-weighted estimation (W.PS) over alternatives across all scenarios. We therefore recommend using survey-weighted propensity score models as the default when survey weights are available, particularly for weighting methods targeting population effects.
An important practical consideration concerns the specification of $e_{\text{fp}}(\mathbf{x})$ under retrospective sampling. While Equation \eqref{eq:efp-retro-ps} reveals a complex nonlinear form, our sensitivity analyses (Web Appendix B.3) demonstrate that under rare sampling conditions (survey weights $>10$, corresponding to sampling rates $<10\%$), a standard unweighted logistic regression provides a robust and computationally feasible approximation for $e_{\text{fp}}(\mathbf{x})$ when combined with survey-weighted estimation of $e_{\text{sp}}(\mathbf{x})$ in augmented estimators. For studies with higher sampling rates (survey weights $<10$), researchers should consider explicitly modeling the nonlinear form through maximum likelihood estimation that incorporates knowledge of the sampling mechanism, though this requires stronger assumptions about the sampling design and increased computational complexity.

Second, augmented estimators are instrumental in reducing bias, enhancing efficiency, and improving coverage in survey observational data, particularly when both the propensity score and outcome models are correctly specified. Our extensive simulation results highlight the strengths of MOM, CVR and WET in different scenarios. WET demonstrates superior robustness under model misspecification and consistently performs well across estimands and overlap conditions. 
CVR shows high efficiency under good overlap and correct models but exhibits sensitivity to outcome model misspecification and diminished reliability in poor overlap scenarios. These findings align with prior literature \cite{Kang2007, Robins2007} that highlights the instability of inverse propensity-based covariate approaches (without survey sampling), particularly under extreme values. Extrapolation issues are exacerbated when the clever covariate incorporates large survey weights, amplifying variability and undermining model performance. 
Reformulating the clever covariate to preserve boundedness properties\cite{Robins2007} could enhance its robustness and practical utility, particularly in scenarios with extreme survey weights or limited overlap, and is an area for future research.

Third, WET consistently emerges as the most robust and efficient estimator across diverse overlap conditions and model misspecifications. 
Compared to CVR and MOM, WET demonstrates superior stability and precision, particularly in poor overlap or when models are misspecified. 
This aligns with Kang and Schafer\cite{Kang2007} and Robins et al.\cite{Robins2007}, who suggest that clever covariates can supplement outcome models but fail to address misspecifications under poor overlap, leading to coverage declines as low as 30\%, as shown in Table \ref{PoorOverlapSimulationResults}.
Moreover, as noted by Robins et al.,\cite{Robins2007} when the CVR is structured with an identity link, it aligns with Targeted Maximum Likelihood Estimation (TMLE) for marginal means but can perform worse than the weighted regression estimator (WET) when both models are misspecified. 
Simulation results and case studies further confirm the robustness of WET, showing lower bias and more reliable coverage in both prospective and retrospective survey designs.
Our case study analyses also demonstrate the capacity of WET to mitigate the influence of extreme survey weights, particularly for PATO. Overall, WET under W.PS can be considered as a recommended approach for causal inference with complex surveys.

The validity of our study’s findings hinges on critical assumptions, including the stable unit treatment value assumption (SUTVA) and treatment assignment ignorability. \cite{rosenbaum1983central} These assumptions are crucial for unbiased causal effect estimation under observational designs but may be violated. Hence, sensitivity analyses \cite{rosenbaum1983assessing}  could provide critical insights into the potential impact of unmeasured confounding for weighting and augmented weighting estimator, and need to be generalized to complex survey data. Second, our analysis assumes that survey weights are known and derived from well-designed surveys. If this assumption fails to hold, causal inferences may be compromised. Insufficient survey designs, including unaddressed non-response biases or inadequate post-stratification, can result in flawed survey weights that exacerbate selection bias or produce extreme values, destabilizing estimators and reducing efficiency. Future research could extend sensitivity analyses to explore the robustness of survey weights under alternative assumptions about the sampling mechanism.
When design weights are unavailable entirely, as in voluntary or nonprobability samples, alternative approaches such as generalized entropy calibration\cite{kwon2025generalizedentropycalibrationanalyzing} offer complementary frameworks for selection bias adjustment.
Third, while our findings focus on binary treatments, they are also readily extensible to multiple treatments. With multiple nominal or ordinal treatments, the generalized propensity scores (GPS) provide a key device for simultaneously balancing covariates across multiple treatment arms, \cite{Li2019b} and are key ingredients in developing weighting and augmented weighting estimators. Future work is needed to extend the current methods to multiple treatments in the context of survey data. Finally, our analysis relied on parametric models for propensity score and outcome estimation, reflecting their widespread use in applied practice. Nevertheless, data-adaptive machine learning approaches offer greater flexibility in capturing nonlinear relationships and complex interactions, and may be particularly advantageous in high-dimensional settings or when parametric assumptions are violated. Extending the proposed AIPW framework to incorporate double machine learning estimators based on data-adaptive working models \cite{chernozhukov2018double} for survey-weighted observational data represents an important and promising direction for future research.


\input{main_archive_Rev1_resubmission.bbl}


\end{document}

%% file: main_archive_Rev1_resubmission.bbl
 \newcommand{\noop}[1]{}